\documentclass[aps,twocolumn,showpacs,floatfix,superscriptaddress]{revtex4-1}
\usepackage[dvips]{graphicx}
\usepackage{dcolumn}
\usepackage{amsmath}    
\usepackage{amssymb}      
\usepackage{multirow}
\usepackage{color}
\usepackage{moreverb}
\usepackage{bm}
\newcommand{\be}{\begin{equation}}
\newcommand{\ee}{\end{equation}}
\newcommand{\bfig}{\begin{figure}}
\newcommand{\efig}{\end{figure}}

\newcommand{\g}{graphene}
\newcommand{\gh}{graphane}

\newcommand{\etal}{{\em et al }}

\begin{document}
\title{Magnetic impurities in graphane with dehydrogenated channels}

\author{Soumyajyoti Haldar}
\affiliation{Department of Physics and Astronomy, Division of Materials Theory, Uppsala University, Box 516, SE-75120 Uppsala, Sweden}

\author{Dilip G. Kanhere}
\affiliation{Department of Physics, Central University of Rajasthan, Bander Sindri Campus, Dist-Ajmer, Rajasthan-305801, India}

\author{Biplab Sanyal}
\email[Corresponding author: ]{Biplab.Sanyal@physics.uu.se}
\affiliation{Department of Physics and Astronomy, Division of Materials Theory, Uppsala University, Box 516, SE-75120 Uppsala, Sweden}

\date{\today} \begin{abstract} We have investigated the electronic and magnetic
    response of a single and a pair of interacting Fe atoms placed in patterned
    dehydrogenated channels in graphane within the framework of density functional
    theory. We have considered two channels -- ``armchair'' and ``zigzag''
    channels. Fully relaxed calculations have been carried out for three different
    widths of the channels. Our calculations reveal that the response to the
    magnetic impurities is very different for these two channels. We have also
    shown that one can stabilize magnetic impurities (Fe in the present case)
    along the channels of bare carbon atoms, giving rise to a magnetic insulator
    or a spin gapless semiconductor. Our calculations with spin-orbit coupling
    shows a large in-plane magnetic anisotropy energy for the case of an armchair
    channel. The magnetic exchange coupling between two Fe atoms placed in the
    semiconducting channel with an armchair edge is very weakly ferromagnetic
    whereas a fairly strong ferromagnetic coupling is observed for reasonable
    separations between Fe atoms in the zigzag edged metallic channel with the
    coupling mediated by the bare carbon atoms. The possibility of realizing an
    ultra-thin device with novel magnetic properties is discussed.

\end{abstract}

\maketitle
\section{\label{sec:intro}Introduction}

In the last few years graphene has generated an intense amount of activities due to its numerous unique properties. \cite{graph1, graph2, graph3} Apart from a very strong multidisciplinary interest in the properties of pure graphene, the scientific community is engaged heavily in the modifications of graphene to realize new and novel effects. There is a tremendous interest in tuning the electronic structure of graphene to open up an energy gap suitable for the applications in electronic devices. One of the ways is by the chemical modification of graphene \cite{sofo:153401, elias} and the other is to realize confined structures in the form of nanoribbons, nanodots etc. \cite{loui,cohen,kim-prl,dai} In this context, a considerable research effort is continuing in attaching hydrogen \cite{boukhvalov,casolo,lebegue,flores,Wu,sahin-apl} or fluorine atoms \cite{nrl-cf,peeters-cf,geim-small,wang-small,zhu-prb,ciraci-cf} to graphene in order to have a stable composite system with a desired energy gap.  

Graphane is a hydrogenated graphene lattice with one hydrogen atom attached to each carbon atom giving rise to an insulating system with sp$^3$ bonds, first predicted by ab initio theory \cite{sofo:153401} and then confirmed by experiments. \cite{elias} The most attractive aspect of hydrogenation in graphene is the opening of a band gap. In a previous paper, we studied various concentrations of hydrogenation on graphene. We showed that a semimetal to metal to insulator transition occurs as a function of hydrogen concentration in graphene and hydrogenation favors clustered configurations leading to the formation of compact islands. \cite{prachi}

One of the results in our earlier work as well as work reported by others \cite{prachi,aksingh,sessi} showed that the patterning of {\gh} via hydrogen desorption, the partial hydrogenation with the simultaneous presence of bare and hydrogenated carbon atoms, is a novel way of modifying the properties of pure {\g}. It turns out that such a pattern  can lead to conducting channels, quantum dots, opening of band gap and finally magnetically coupled graphene/graphane interfaces. These interesting properties of {\g}/{\gh} interface structures have attracted a large number of researchers. \cite{aksingh,sessi,ylu,balog,zm-ao} Recent studies show that the energy band gaps of both armchair and zigzag {\gh} nanoribbons decrease as the nanoribbons become wider. \cite{ylu,yli} A variety of other interesting properties such as thermal stability, interface engineering, thermal conductivity, etc. have also been investigated for these hybrid systems. \cite{zm-ao,kowsary} Recently zigzag graphene nanoribbons have been experimentally synthesized by selectively removing hydrogens from epitaxial graphane. \cite{ijas} Thus patterning offers a definite possibility of generating stable nanoribbons in the form of patterned {\g} nanoribbons. 

Graphene being a potential candidate for spintronic device has led to a number of theoretical and experimental studies on the determination of the magnetic properties of graphene. \cite{yazev-rev} The existence of a spin-polarized edge state has been predicted by a number of theoretical works, when graphene sheet is cut to have parallel zigzag edges (called zigzag graphene nanoribbon or ZGNR). \cite{nakada,loui,hofman,girit,koski-prb,koski-prl,cohen,mag2,bhandary} However, armchair graphene nanoribbon (AGNR) does not show such a spin-polarized state. The spin-polarized edge state in ZGNR is predicted to be highly dependent on the edge geometry and may not be robust. \cite{jens}. Recently it was shown that transition metal (TM) termination at the edge transforms semiconducting ZGNR to a metallic one. \cite{haiping1} It was also shown that Fe terminated ZGNR shows an anti-ferromagnetic (AFM) coupling between the two edges of the nanoribbon. \cite{haiping2} Furthermore, it has been showed recently that the magnetic properties of transition metal doped ZGNRs are more robust than those moments arising due to the edge geometry. \cite{power} The most attractive feature of patterning graphene with partial hydrogenation is the similarity of creating bare carbon channels having similar edge state properties as those of graphene nanoribbons. Recently edge states of {\g}/{\gh} interface has been studied using tight-binding approximation and it was shown that edge state of an interface oriented along a zigzag direction enhances the effects related to spin-orbit interaction. \cite{loss} A recent study of zigzag {\g} nanoribbon patterned on {\gh} using spin polarized \emph{ab initio} calculations showed that the electronic and magnetic properties of the {\g}/{\gh} superlattice strongly depends on the degree of hydrogenation at the interfaces between the two materials. \cite{nieves} 

Quite clearly the nature of electronic states in these two channels are different. The electrons in armchair channel are localized and non magnetic. However the electrons in zigzag channel are delocalized. A weak anti-ferromagnetic coupling across the edges exists which decreases as the width increases. It will be interesting to investigate the effect of magnetic impurities in both these channels. Therefore we have carried out spin-polarized \emph{ab initio} density functional theory calculations (DFT) to study the electronic and magnetic response properties of one and two magnetic impurities placed in these channels. We have considered three different widths of the channels along with two to three possible Fe-Fe distances. We have also investigated the nature of the magnetic exchange coupling between two Fe impurities.

There is another aspect that needs attention namely the effect of spin orbit coupling. One may expect an increase in the orbital moment of nanostructures compared to that of the bulk due to reduced symmetry. An experimental determination of magnetic moments includes both spin and orbital moments. Another point of relevance is the easy axis of magnetization. Therefore, we have calculated the orbital magnetic moments including spin-orbit coupling (SOC) in the Hamiltonian and have also discussed the strength of magnetic anisotropy along with the easy axis of magnetization. It is noteworthy that one of us has shown recently \cite{fep} that the magnetic anisotropy energy (MAE) of an Fe atom of Fe-porphyrin molecule deposited on a defected graphene lattice can be manipulated by straining the graphene lattice. In that case, the predicted MAE is between 0.4 and 2 meV. So, in the present case, it is interesting to study the effect of SOC induced changes in magnetism of the Fe impurities.

The plan of the paper is as follows. In the next section we will describe in some detail our calculations followed by the results in Sec.~\ref{sec:res}. We will briefly recapitulate some of the important features of graphene/graphane interface. In Sub Sec.~\ref{subsec:1fe}, we will discuss the effect of adsorbing a single Fe atom on both types of channels for different widths followed by the results on spin-orbit coupling induced orbital magnetic moments and magnetic anisotropy energies (MAEs) for a particular case in Sub Sec.~\ref{subsec:mae}. Then we discuss the results on the magnetic interactions between two Fe atoms in the channels. The conclusions are given in Sec.~\ref{sec:concs}. 
 
\section{\label{sec:compute}Computational Details}

We have used plane wave based density functional theory (DFT) as implemented in VASP code \cite{vasp,vasp1} for the calculations. The projector augmented wave (PAW) method \cite{Blo,Blo1} within the generalized gradient approximation as given by Perdew, Burke and Ernzerhof \cite{pbe,pbe2} has been used for the exchange-correlation potential.  The electronic wave functions were expanded using plane waves up to a kinetic energy of 450 eV. The spin-polarization approach was implemented in all calculations. For electron smearing, Fermi smearing was used and was kept fixed at 0.05 eV. The energy and the Hellman-Feynman force thresholds were kept at $10^{-5}$ eV and 0.005 eV/\AA~ respectively. We considered the chair conformer configuration of monolayer {\gh}, as described by Sofo \etal. \cite{sofo:153401}  The pristine system was modeled using a supercell approach.  The separation between the graphane layers in two consecutive unit cells in the perpendicular direction is kept as large as 15 {\AA} to avoid the interactions between the two {\gh} planes. For geometry optimization, a 5 $\times$ 5 $\times$ 1 {\bf $k$} grid was used whereas a 11 $\times$ 11 $\times$ 1 {\bf $k$} grid was used to calculate the Density of States (DOS) in the Monkhorst-Pack scheme. All atomic positions were allowed to relax where only the forces between the atoms are minimized and the volume of the elemental cell is kept constant. This volume is equal to the volume of an equivalent graphane system. This is similar to the experimental situation where graphene nanoribbons are designed in a graphane sheet with fixed boundary. We have tested the influence of the cell relaxation and have found that the lattice parameter is decreased by less than 1$\%$, which is insignificant for our present discussion. The use of a fixed boundary with the lattice parameter of graphane was justified in an earlier study \cite{nieves}, where it was argued that the relaxations of the internal coordinates is sufficient to describe the electronic structure and magnetism at the graphene/graphane interface. For the calculations of orbital moments and magnetic anisotropies, the energy convergence was set to $10^{-6}$ eV. 

The binding energy $E_b$ of Fe, is defined as
\begin{equation*}
E_{b} = [N*E(Fe) + E(Graphane)] - E(Fe_N+Graphane) 
\end{equation*} where
\begin{itemize}
\item[-] $E(Fe_N+Graphane)$ is the electronic energy for equilibrated, optimized Fe$_N$+{\gh} system.
\item[-] $E(Graphane)$ is the electronic energy for the optimized {\gh} geometry. 
\item[-] $E(Fe)$ is the electronic energy for the isolated Fe atom placed in a cubic supercell of length 10~\AA. Only the $\Gamma$ point of the Brillouin zone was sampled in this case. 
\item[-] $N$ is the number of Fe atoms.
\end{itemize}

\section{\label{sec:res}Results and Discussions}

\begin{figure}[!hbp]
\begin{center}
\includegraphics[scale=0.20]{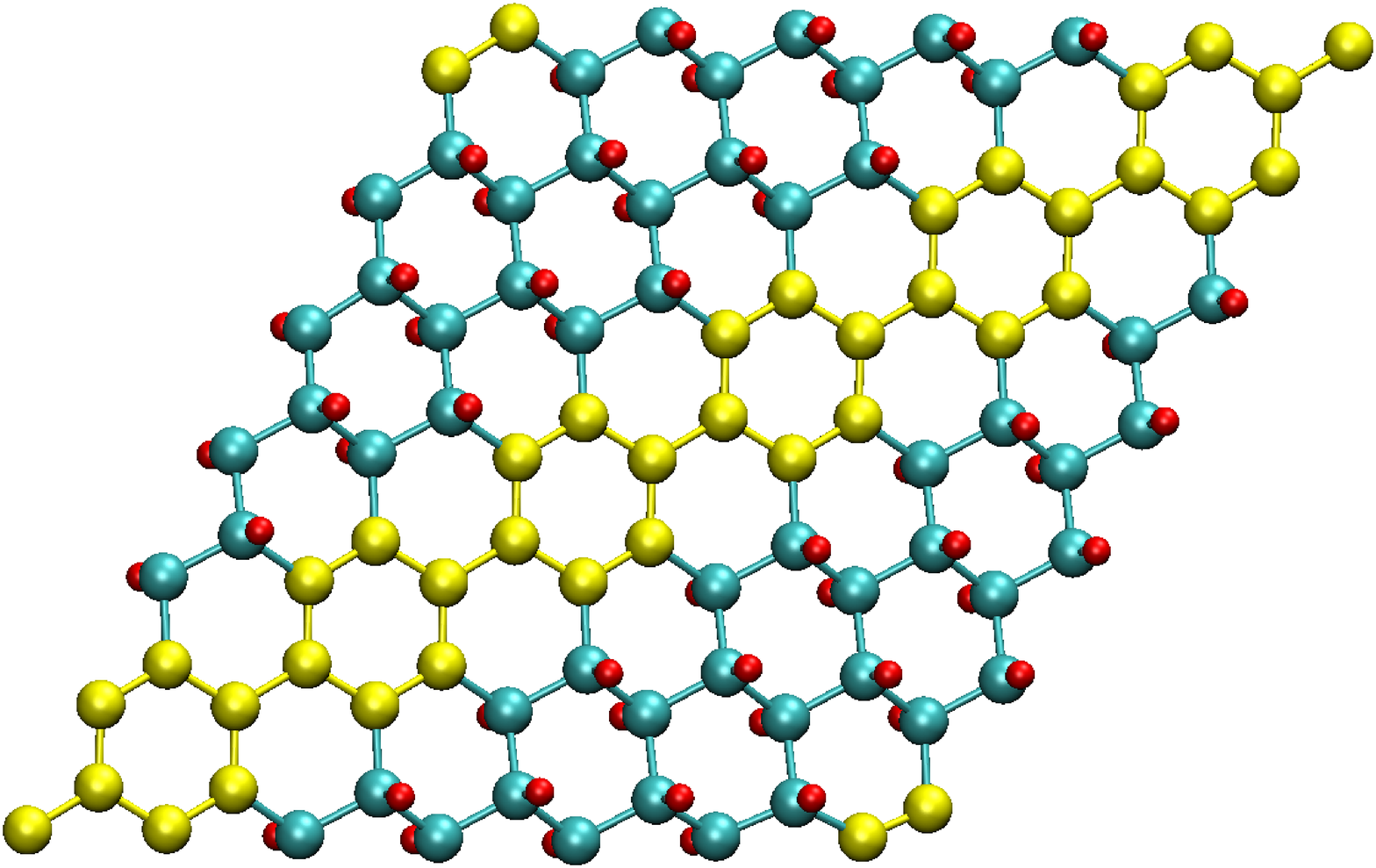}\\~\\
(a)``armchair'' channel\\~\\
\includegraphics[scale=0.20]{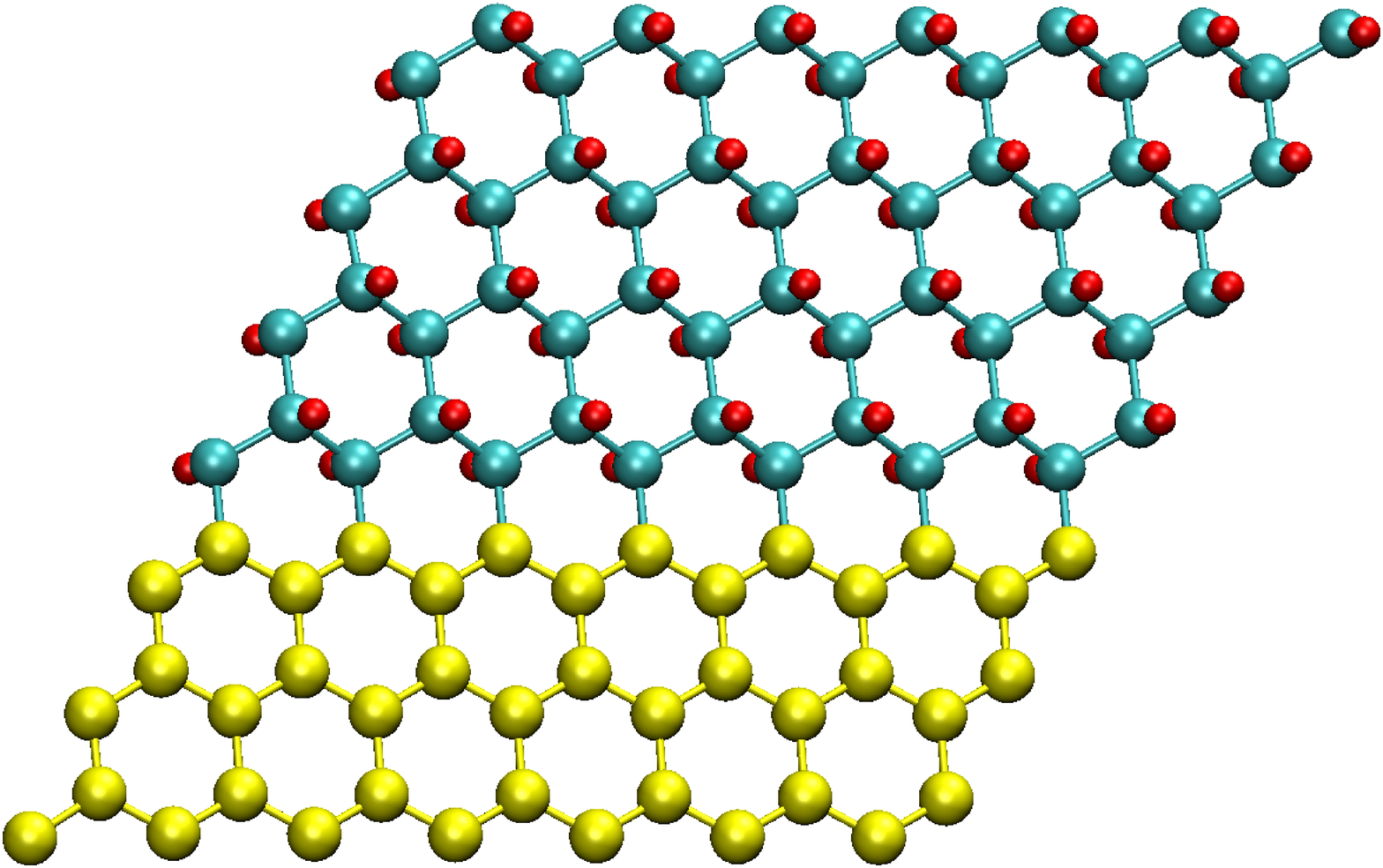}\\~\\
(b)``zigzag'' channel\\~\\
\end{center}
\caption{\label{fig:geom} (Color online) Decoration of hydrogen (a) along the diagonal of the unit cell leading to an ``armchair'' channel and (b) along the edge of the unit cell leading to a ``zigzag'' channel. In the figure, yellow (light shaded in print) balls are bare carbon atoms, turquoise (dark shades in print) balls are hydrogenated carbon atoms and red (small dark in print) balls are H atoms. For both armchair and zigzag channels, the nanoribbons are 3-rows wide.}
\end{figure}

We briefly recapitulate the interesting result of two types of {\g}/{\gh} channels. These two channels are schematically shown in Fig.~\ref{fig:geom}. Figures~\ref{fig:geom}(a) and~\ref{fig:geom}(b) show ``armchair" and ``zigzag" channels created by removing hydrogen atoms along the diagonal and the edges of the supercell respectively for a specific width of 3 rows. While the ``armchair channel" is non magnetic and has a width dependent bandgap, the ``zigzag channel" interfaces are magnetic. The lowest energy structure is antiferromagnetically coupled across the edge but as the width increases, the FM and AFM coupled edges become almost degenerate e.g., FM-AFM energy difference decreases to 3.63 meV/edge C atom from 6.57 meV/edge C atom as the width of the zigzag channel increases from 3-rows to 7-rows.  Beyond a certain width, the exchange coupling between the edges becomes zero with C moments along each edge ferromagnetically aligned.

\subsection{\label{subsec:1fe}Single Fe atom on bare channels}
We begin by presenting the results of a single Fe atom placed on three different widths of these channels. It is known that for pristine {\g}, the energetically favorable position of an Fe atom is at the hollow site of the hexagonal ring with an Fe to carbon distance of 2.1~\AA. \cite{zanella,valencia,chan-cohen} Our calculations show that the minimum energy position of Fe is dependent on the type of channel. For the armchair channel the lowest energy position of Fe is at the hollow site equidistant from the interface for all the widths. However in the zigzag channel, the minimum energy position of Fe is when it is at the hollow site of bare carbon atoms nearest to the interface. This is because of the anti-ferromagnetic nature of the edge coupling. For the 3-rows  channel, the central row is nearest to both the interfaces and Fe atom prefers to sit there. However as the width increases, the Fe atom shifts towards the interface. The binding energy of the structure where Fe is in the interface is higher by 0.31 eV/Fe atom and 0.36 eV/Fe atom respectively for the 5-rows and the 7-rows channel than the Fe atom at the center of the channel. The position of Fe is slightly asymmetric w.r.t.\ the surrounding hexagon due to the stretching of underlying carbon bonds and the Fe to C distance varies from 2.06 {\AA} to 2.18 {\AA}. 

\begin{table}[tb]
\begin{center}
\caption{\label{tab1}Width dependent energies and magnetic moments of the channel systems with a single Fe impurity. $E_b$, $\mu_{\text{total}}$ and $\mu_{\text{Fe}}$ denote the binding energy of Fe, total magnetic moment of the system and magnetic moment on Fe site respectively.\\}
\begin{tabular}{|c|c|c|c|c|} \hline 
\multicolumn{5}{|c|}{Single Fe atom} \\ \hline
\multirow{2}{*}{Channel Type} & \multirow{2}{*}{Channel Width} & $E_b$ & $\mu_{\text{total}}$ & $\mu_{\text{Fe}}$ \\
& & (eV) & ($\mu_B$) & ($\mu_B$) \\ \hline
\multirow{3}{*}{Armchair} & 3-rows & 1.6 & 2.0 & 1.99\\ \cline{2-5} 
& 5-rows & 1.10 & 2.06 & 2.10\\ \cline{2-5}
& 7-rows & 0.80 & 2.00 & 1.98\\ \hline 
\multirow{3}{*}{Zigzag} & 3-rows & 1.38 & 2.1 & 2.5\\ \cline{2-5} 
& 5-rows & 1.42 & 2.01 & 2.55\\ \cline{2-5}
& 7-rows & 1.45 & 2.01 & 2.56\\ \hline 
\end{tabular}
\end{center}
\end{table}

\begin{figure}[tb]
\begin{center}
\includegraphics[scale=0.3]{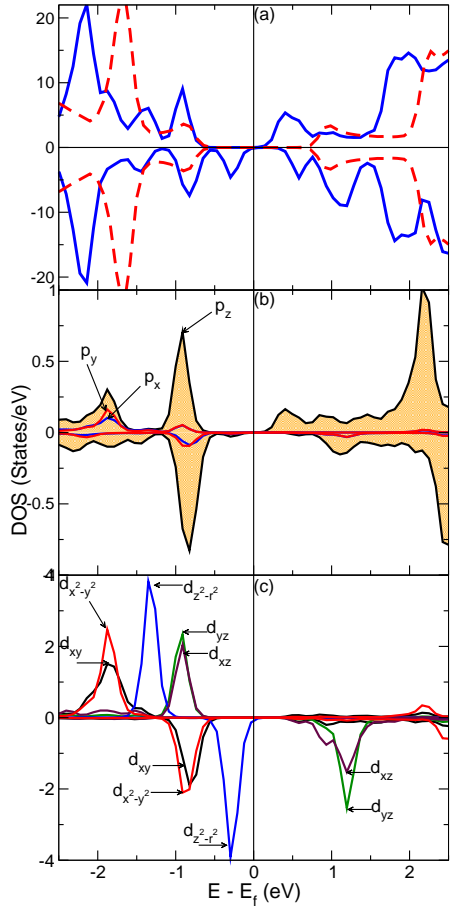}
\end{center}
\caption{\label{fig:diag-pdos}(Color online) Spin polarized DOS for single Fe atom placed on the 3-rows ``armchair'' channel. (a) Total DOS with (blue solid line) and without (red dashed line) Fe. (b) Site and angular momentum projected DOS for Fe surrounding six C atoms. The orange (in print, light grey) shaded region shows the $p_z$ component. (c) Site and angular momentum projected DOS ($d$ components only).}
\end{figure}

\begin{figure}[tb]
\begin{center}
\includegraphics[scale=0.3]{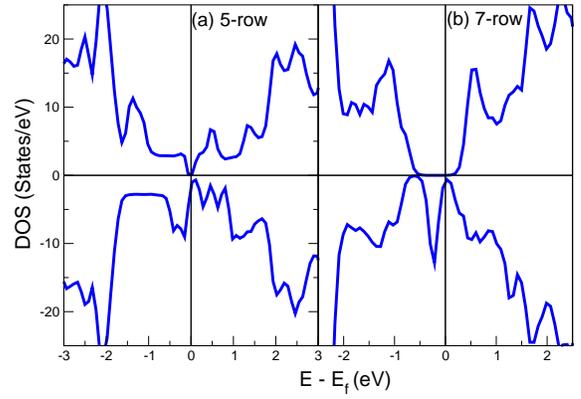}
\end{center}
\caption{\label{fig:diag-dos-57} (Color online)Total DOS for 5-rows and 7-rows ``armchair'' channel with a single Fe impurity.}
\end{figure}

Now we will discuss the energetics and electronic structure of a single Fe atom placed in both the channels for all widths. Table~\ref{tab1} shows the energies and magnetic moments of the Fe atom placed in these channel systems. From the table, it can be seen that as the channel width increases, the binding energy of Fe atom for the armchair channel decreases while for the zigzag channel it remains more or less constant. We also note that the total magnetic moments for both the channels for all the widths are $\sim$ 2.0 $\mu_B$, consistent with the value for magnetic moment of a single Fe atom on {\g}. \cite{valencia} However, the on-site local magnetic moment on the Fe atom is $\sim$ 0.5 $\mu_B$ higher in the zigzag channel (see the discussion next). It is also observed that the binding energy of Fe in all three different width zigzag channel structures are about 0.2 eV higher than the same in pure {\g}. \cite{valencia} Thus the mixed $sp^2-sp^3$ character increases the binding energy of Fe in the zigzag channels.

\begin{figure}[tb]
\begin{center}
\includegraphics[scale=0.18]{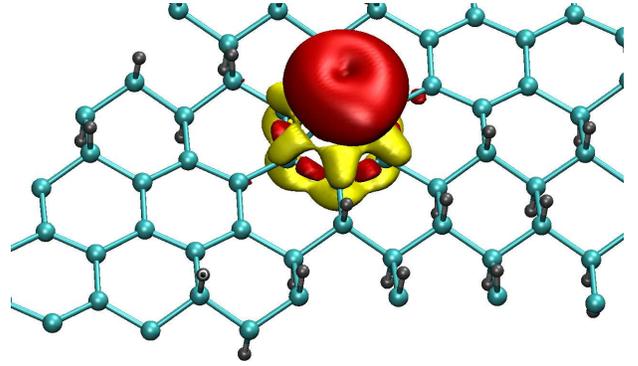}
\end{center}
\caption{\label{fig:diag-spind} (Color online) Spin density plot of a single Fe atom on the 3-rows ``armchair'' channel. Red (in print dark shade) is positive spin density and yellow (in print light shade) is negative spin density.}
\end{figure}

We further analyze the result by examining the DOS and spin densities of both the channels. In Fig.~\ref{fig:diag-pdos} we show spin polarized total DOS with (blue solid line) and without (red dashed line) Fe (fig.~\ref{fig:diag-pdos}(a)), site and angular momentum projected DOS for Fe and Fe near-neighbor C atoms for 3-rows armchair channel (fig.~\ref{fig:diag-pdos}(b-c)). As can be seen from the DOS, there are induced states due to spin-down electrons appearing in the gap below the Fermi energy. These states are due to Fe $d$ spin-down electrons. As a consequence, the gap reduces significantly. This gap is in between spin-up and spin-down electrons. Analysis of total DOS for 5-rows and 7-rows armchair channel (shown in fig.~\ref{fig:diag-dos-57}) indicates that this feature is also present as the width of the channel increases from 3 to 7 rows. However the value of the gap depends on the width of the channel.

The nature of spin densities (shown in fig.~\ref{fig:diag-spind}) can be analyzed with the help of site and angular momentum decomposed density of states (henceforth called as PDOS) (shown in fig.~\ref{fig:diag-pdos}(b-c)). The spin density plot shows that the majority of spin-up density is localized only on Fe atom and Fe is seen to induce the spin-down density on the surrounding six C atoms. There is a large localized magnetic moment of the order of 2 $\mu_B$ on Fe (see table~\ref{tab1}) and a very small negative magnetic moment on the surrounding six C atoms ($\sim$ -0.02 $\mu_B$/atom). Interestingly there is an induced up spin density in between the carbon atoms of Fe surrounding ring. Analysis of PDOS shows that the Fe $d_{xz}$ and $d_{yz}$ of up spin channel interact with the $\pi$ bonded system formed by $p_z$ orbitals of the surrounding carbon atoms (shown in fig.~\ref{fig:diag-pdos}(b)). This induces polarization on the ring so as to generate negative moments on the C sites and positive moment in between the C atoms (see spin density in fig.~\ref{fig:diag-spind}). It is to be noted that the magnetic moment of Fe is reduced from $4\mu_B$ for isolated atom to $2\mu_B$ for the system. The analysis of angular momentum decomposed DOS of Fe (fig.~\ref{fig:diag-pdos}(c)) shows a strong hybridization of Fe $3d$ states with the $p_z$ orbitals of the C atoms. The DOS shows that all five spin-up $3d$ states and three spin-down $3d$ states of Fe are occupied. Both spin-up and spin-down $4s$ states are above the Fermi energy indicating that two electrons are transferred from atomic Fe $4s$ states to spin-down $3d$ states of Fe-channel system. This causes the reduction of magnetic moment on Fe site which is consistent with the earlier result. \cite{chan-cohen}. For larger widths ( 5-rows and 7-rows), we have also analyzed the site and angular momentum projected DOS (figure not shown) and the behavior of the spin densities are found to be similar in nature.

\begin{figure}[!htbp]
\begin{center}
\includegraphics[scale=0.3]{fig5a}\\
(I)\\~\\
\includegraphics[scale=0.18]{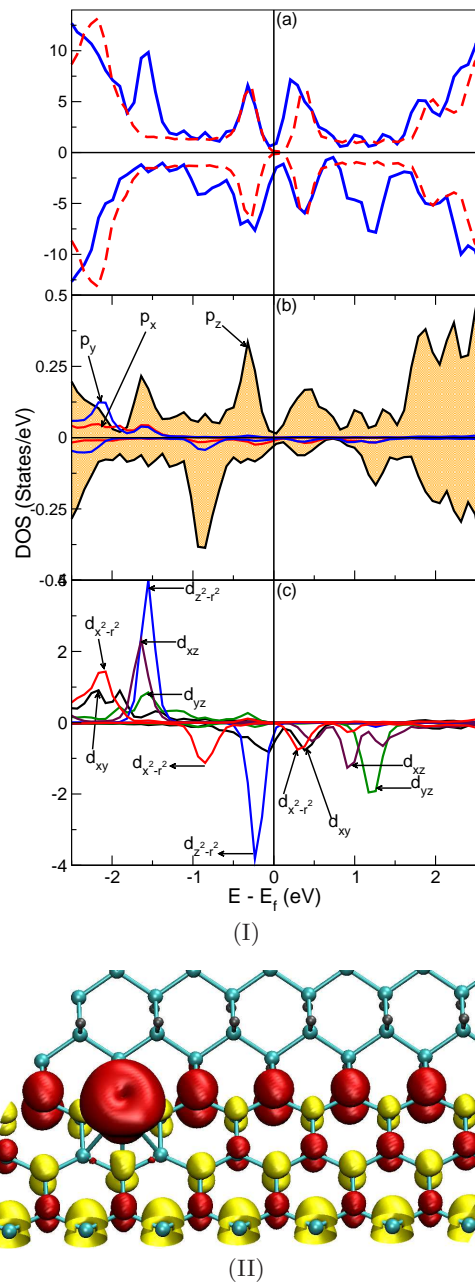}\\
(II)
\end{center}
\caption{\label{fig:fe1-edge}(Color online) (I) Spin polarized DOS for a single Fe atom placed on the 3-rows ``zigzag'' channel. (a) Total DOS with (blue solid line) and without (red dashed line) Fe. (b) Site and angular momentum projected DOS for Fe surrounding six C atoms. The orange (in print light grey) shaded region shows the $p_z$ component. (c) Site and angular momentum projected DOS ($d$ components only). (II) Spin density plot of a single Fe atom on the 3-rows ``zigzag'' channel. Red (in print dark shade) is positive spin density and yellow (in print light shade) is negative spin density.}
\end{figure}

Fig.~\ref{fig:fe1-edge}-(I) shows spin polarized total and site projected DOS for the 3-rows zigzag channel. The red dashed line in the total DOS is for a bare zigzag channel. In contrast to the armchair channel, the DOS for the edge channel (fig.~\ref{fig:fe1-edge}(a)) shows a finite DOS near the Fermi energy. This is mainly because of spin-down electrons from the Fe site and bare C sites with peaks below and above the Fermi energy. The detailed examination of site and angular momentum projected DOS shows that the down spin peak just below the Fermi energy comes from both Fe $d_{z^2-r^2}$ orbitals and the $p_z$ orbitals of the interface bare carbon atoms. The contribution in up spin peak just below the Fermi energy is from delocalized $p_z$ orbitals of edge bare C atoms as well as a small contribution coming from $\pi$ bonded C atoms surrounding the Fe atom. 

Our analysis also show that the effect of magnetic impurity is felt up to 4th nearest neighbor interface C atoms. It reduces the on-site magnetic moments on edge C atoms and the maximum reduction is $\sim$ 15\% on the nearest site. This is a sharp contrast compared to the armchair channel where the Fe impurity affects only up to 1st nearest neighbor. Another interesting difference is in the induced spin density on the hexagonal C ring nearest to Fe. In the present case this ring is not anti-ferromagnetically ordered. Only three C atoms (belonging to same sub lattice) show significant down spin density (see in fig.~\ref{fig:fe1-edge}-(II)). This is understandable from the fact that the delocalized electron density present in the channel interacts with the Fe atom. It reduces the amount of charge transfer from Fe $4s$ states to Fe $3d$ states, resulting in an increment of local magnetic moment on Fe to $2.5\mu_B$. This also leads to a somewhat long ranged perturbation in the underlying C lattice.

\subsubsection{\label{subsec:mae} Role of spin-orbit coupling}
As discussed in the introduction, we have included SOC in the Hamiltonian to calculate the orbital moments and magnetic anisotropy energies for the case of 3-rows wide armchair and zigzag channels with a single Fe atom. The  calculated orbital moment of the Fe atom in the armchair channel is 0.09 $\mu_{B}$, which is enhanced compared to the calculated value for bulk Fe ($\sim$ 0.05 $\mu$ without orbital polarization) in the bcc phase. This is expected due to the lowering of symmetry in the present case. However, the enhancement is not dramatic due to the strong crystal field produced by the C atoms in the hexagon to which Fe is chemically bonded. The calculated easy axis of magnetization is in-plane with MAE equal to 19 meV/Fe. This strong magnetic anisotropy should hold the Fe moment in the plane of the substrate. The orbital moment along the 001 direction (hard axis) is 0.02 $\mu_{B}$, consistent with the fact that the orbital moment is reduced along the hard axis. We have observed an interesting effect in the case of the zigzag channel. In the presence of SOC, some of the C atoms have spin moments of the order of 0.1 $\mu_{B}$. Therefore the total moment of the unit cell in much increased (4.5 $\mu_{B}$) compared to the case without SOC. Although the Fe spin moment stays more or less the same, the additive contribution from C atoms gives rise to this enhancement of the total moment. Also, in the case of zigzag channel, the MAE is smaller ($\sim$ 4 meV/Fe) with an in-plane easy axis. The in-plane orbital moment amounts to 0.1 $\mu_{B}$, which is slightly larger than the out-of-plane moment (0.08 $\mu_{B}$) The topic on the magnetic anisotropy of adatoms on graphene and the tuning of MAE with an electric field has been discussed very recently. \cite{stefan} 

\subsection{\label{subsec:2fe} Magnetic Interaction between two Fe atoms}
Now we will discuss the case of two interacting Fe impurities. As expected, it turns out that the response of the channel electrons are different for the armchair and zigzag channels. Table~\ref{tab2} shows the binding energy of Fe atoms, total magnetic moment, on-site magnetic moment on Fe atoms and exchange energies (wherever possible). For armchair channel we have considered two possible Fe-Fe distances for all the channel widths. In this channel the two magnetic impurities (Fe) are weakly interacting as evident from the weak variation in the binding energy of Fe atoms as a function of distance coupled with the fact that the total magnetic moment remains almost unchanged. Consequently the exchange energy ($E^{ex}$) is also negligible. The spin densities are also similar to the isolated single Fe atom in the armchair channel (additive in nature). Our calculations show that these features are also present for the larger widths. 

\begin{table}[!hbp]
\begin{center}
\caption{\label{tab2}Width dependent energies and magnetic moments of the channel systems with a pair of Fe atoms with possible Fe-Fe distances. The distances between two Fe atoms ($d$) are in Angstrom, binding energies of Fe in electron volt ($E_b$). Total magnetic moment of the systems ($\mu_{\text{total}}$) and magnetic moment on the Fe atoms ($\mu_{\text{Fe}}$) are in the units of Bohr-magneton and the exchange energies ($E^{ex}=E^{FM}-E^{AFM}$) are in electron volt\\}
\begin{tabular}{|c|c|c|c|c|c|c|} \hline 
\multicolumn{7}{|c|}{Two Fe atoms in the channel} \\ \hline 
\multirow{2}{*}{Type} & \multirow{2}{*}{Width} & $d_{\text{Fe}}$ & $E_b$ & $\mu_{\text{total}}$ & $\mu_{\text{Fe}}$ & $E^{ex}$\\
& & (\AA) & (eV) & ($\mu_B$) & ($\mu_B$) & (eV)\\ \hline
\multirow{6}{*}{armchair} & \multirow{2}{*}{3-rows} & 4.38 & 3.13 & 4.02 & 1.98/2.01 & 0.009 \\ \cline{3-7}
& & 8.77 & 3.18 & 4.01 & 1.99/1.99 & 0.001 \\ \cline{2-7}
& \multirow{2}{*}{5-rows} & 4.38 & 2.25 & 4.11 & 2.10/2.10 & 0.018 \\ \cline{3-7} 
& & 8.77 & 2.19 & 4.08 & 2.11/2.12 & 0.008 \\ \cline{2-7}
& \multirow{2}{*}{7-rows} & 4.38 & 1.75 & 4.06 & 2.04/2.03 & 0.014 \\ \cline{3-7} 
& & 8.77 & 1.64 & 4.01 & 2.01/2.01 & 0.016 \\ \hline
\multirow{9}{*}{zigzag} & \multirow{3}{*}{3-rows} & 2.17 & 4.12 & 5.96 & 2.98/2.98 & 0.457 \\ \cline{3-7}
& & 5.08 & 2.67 & 7.0 & 2.59/2.6 & 0.058 \\ \cline{3-7}
& & 7.60 & 2.67 & 4.26 & 2.42/2.43 & 0.038 \\ \cline{2-7}
& \multirow{3}{*}{5-rows} & 2.17 & 4.09 & 5.89 & 3.01/3.01 & 0.834 \\ \cline{3-7} 
& & 5.08 & 2.73 & 4.14 & 2.53/2.53 & NA \\ \cline{3-7}
& & 7.60 & 2.80 & 4.06 & 2.51/2.51 & NA \\ \cline{2-7}
& \multirow{3}{*}{7-rows} & 2.17 & 4.10 & 5.82 & 3.01/3.01 & 0.795 \\ \cline{3-7} 
& & 5.08 & 2.80 & 4.07 & 2.53/2.54 & NA \\ \cline{3-7}
& & 7.60 & 2.87 & 4.02 & 2.52/2.52 & NA \\ \hline
\end{tabular}
\end{center} 
\end{table}

\begin{figure}[!hbp]
\begin{center}
\includegraphics[scale=0.4]{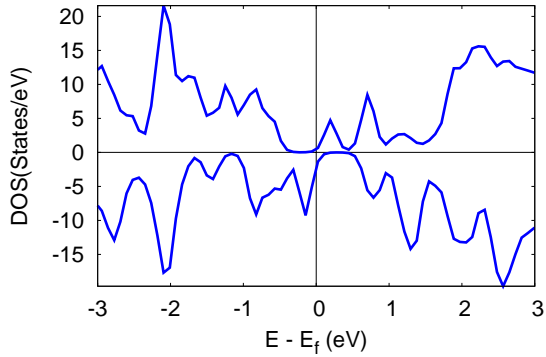} \\
(a) Fe2 DOS - 3-rows (4.38\AA) \\~\\
\includegraphics[scale=0.4]{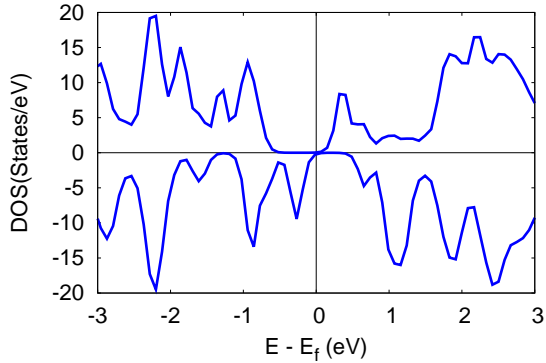} \\
(b) Fe2 DOS - 3-rows (8.77\AA)\\
\end{center}
\caption{\label{fig:2fe-3row-dos}(Color online) Spin polarized total DOS plots for 2 Fe atoms in the 3-rows armchair channel. We've plotted the graphs within -3 eV to 3 eV to bring out the feature of DOS near the Fermi level. The numbers in the parenthesis are the distances between the two Fe atoms.}
\end{figure}

However, the examination of DOS shows some interesting features. The DOS of two Fe atoms in the 3-rows armchair channel (fig.~\ref{fig:2fe-3row-dos}(a) and fig.~\ref{fig:2fe-3row-dos}(b)) shows that both Fe atoms induce states below and above the Fermi energy. It is interesting to note that the DOS shows a gap just below the Fermi energy in the spin-up channel and just above the Fermi energy in the spin-down channel. The observation of DOS for small Fe-Fe distance (see fig.~\ref{fig:2fe-3row-dos}(a)) suggests the possibility of a spin gapless semiconductor material (SGS). It can be seen that there are energy gaps of 0.56 eV and 0.45 eV in the up and down spin channels respectively. However, the cross spin gap between down spin channel and up spin channel is very small. Therefore, at the Fermi energy, an electron has to flip its spin to go to the conduction band from the valence band. The resulting transmission is 100\% spin polarized. Recently Wang has discussed such a possibility in SGSs. \cite{wang-sgs} It was further noted that SGS can be obtained by doping magnetic impurities to semiconductor. In the present case, a similar situation occurs. As the distance between two Fe atom increases, this feature can also be seen there (see fig.~\ref{fig:2fe-3row-dos}(b)). However, as the channel width increases, this feature weakens. Thus there is a possibility of SGS for a narrow strip of armchair channel. 

In contrast to the armchair channel, the binding energy as well as the magnetic moment are distance dependent in the zigzag channel. For the smallest Fe-Fe distance for all the widths, the two Fe atoms forms a dimer.  It is evident from high values of the binding energies as well as the total magnetic moment of the system (see table~\ref{tab2}). For comparison we note that the total magnetic moment of Fe-Fe dimer is 5.94 $\mu_B$. The on-site magnetic moments on Fe atoms are also very high ($\sim$ 3 $\mu_B$ on each Fe atom). 
\begin{figure}[!htbp]
\begin{center}
\includegraphics[scale=0.15]{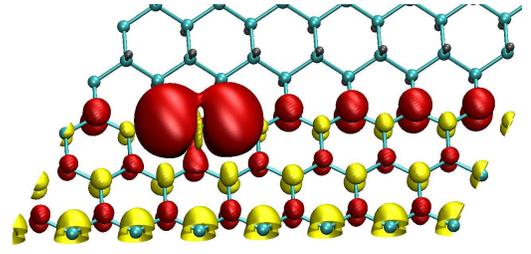}\\~\\
(a) Fe2 Spin 3-rows (2.17\AA)\\~\\
\includegraphics[scale=0.15]{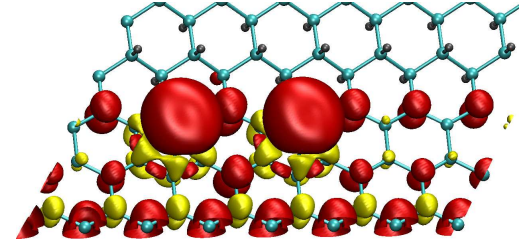} \\~\\
(b) Fe2 Spin 3-rows (5.08\AA)\\~\\
\includegraphics[scale=0.15]{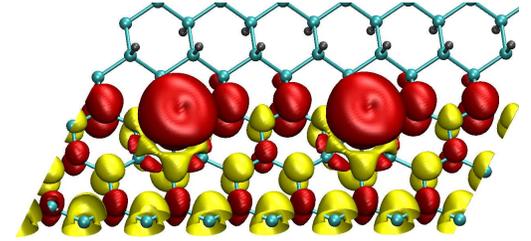} \\~\\
(c) Fe2 Spin 3-rows (7.60\AA)\\~\\
\end{center}
\caption{\label{fig:all-spin} (Color online) Spin density plots. Red (in print
dark shade) is positive spin density and yellow (in print light shade) is
negative spin density.}
\end{figure}

Table~\ref{tab2} also indicates that there is a significant variation in total magnetic moment for the Fe-Fe distance of 5.08 \AA~for all the widths. Although the individual magnetic moment on Fe atoms are similar and of the order of 2.5 $\mu_B$, the total magnetic moment reduces from 7.0 $\mu_B$ (3-rows) to 4.14 $\mu_B$ and 4.07 $\mu_B$ for the 5-rows and the 7-rows channels respectively. In this case the specific position of 2nd Fe atom induces negative spin densities on the surrounding C atoms which in turns affect the spin densities of both edge row C atoms (see fig.~\ref{fig:all-spin}). As a consequence, the channel becomes ferromagnetically coupled across the edges and develops a large total magnetic moment. For the larger widths, as the Fe atoms are placed nearest to one edge, the interaction between the Fe atoms and the other edge is much weakened. Therefore the edges remain weakly anti-ferromagnetically coupled reducing the total magnetic moment.

As expected, the exchange energy $E^{ex}$ is stronger for the smallest Fe-Fe distance (2.17\AA) for all the widths (see table.~\ref{tab2}). For 3-rows width, as the Fe-Fe distance increases, the exchange energy value decreases to 0.038 eV. Thus there is a stable ferromagnetic state for all the Fe-Fe distance with possible transition temperature $\sim$ 350 K. However it was not possible to estimate $E^{ex}$ in other widths because we could not get properly converged anti-ferromagnetically state for longer Fe-Fe distances. 

From the above discussions, it is evident that for the zigzag channel, the magnetic atoms interact strongly. Thus one can envisage to achieve a magnetic graphane lattice by depositing magnetic atoms (Fe in the present case) in the metallic channel of bare carbon atoms. Indeed our preliminary calculations shows that it is possible to obtain stable linear Fe chain by placing Fe atoms nearest to the interface of the zigzag channel. The total magnetic moment and the binding energy was found to be 2.42 $\mu_B$/Fe atom and 2.14 eV/Fe atom respectively. It may be possible to form novel magnetic nanostructures (chains, islands etc.) with the help of scanning tunneling microscopic tips and study magnetism as a function of the width of the bare carbon channels. Therefore, an ultrathin device with novel functionalities may be achieved.

\section{\label{sec:concs}Conclusion}
In conclusion, our detailed density functional investigations have revealed some novel electronic and magnetic response of single and a pair of Fe atoms placed in patterned dehydrogenated channels in graphane. Our work shows that the response of the two channels -- the ``armchair'' and the ``zigzag'' channels are different. In the armchair channel, the magnetic response of the single Fe atom is localized in nature while for the zigzag channel it is fairly long ranged. The easy axis of magnetization is in the plane of graphene with a large magnetic anisotropy energy of 19 meV/Fe atom in the case of the armchair channel. As compared to armchair channel, the binding energies and the total magnetic moments with two Fe impurities are distance dependent in the zigzag channel suggesting that the pair of Fe atoms are rather strongly interacting in this channel. The magnetic coupling between a pair of Fe atoms is very weakly ferromagnetic in the semiconducting armchair channel whereas a relatively stronger ferromagnetic coupling is observed for the closest separation between the Fe atoms in the delocalized zigzag channel. Our result for the armchair channel with two Fe atoms shows a possibility of having a novel spin gapless semiconductor material by doping appropriate magnetic impurities in partially hydrogenated graphene of a particular width. The magnetic interaction between the impurities in partially hydrogenated {\g} with conducting channels shows an oscillating ferromagnetic and anti-ferromagnetic coupling across the edge of the channel for a particular width of zigzag channel. The possibility of realizing an ultrathin device having novel properties is discussed.

\section*{Acknowledgement} 

SH would like to acknowledge Indo-Swiss grant for financial support (No: INT/SWISS/P-17/2009). BS acknowledges financial support from Swedish Research Council, Carl Tryggers Foundation and KOF grant of Uppsala University. SH and DGK acknowledge the Swedish Research Links programme for their visits to Uppsala university. We thank SNIC-UPPMAX and SNIC-HPC2N computing centers under Swedish National Infrastructure for Computing (SNIC) and CDAC-Pune for granting computer time. Some of the figures are generated by using VMD \cite{vmd} and XCrySDen \cite{xcrysden}.


\begin{thebibliography}{58}%
\makeatletter
\providecommand \@ifxundefined [1]{%
 \@ifx{#1\undefined}
}%
\providecommand \@ifnum [1]{%
 \ifnum #1\expandafter \@firstoftwo
 \else \expandafter \@secondoftwo
 \fi
}%
\providecommand \@ifx [1]{%
 \ifx #1\expandafter \@firstoftwo
 \else \expandafter \@secondoftwo
 \fi
}%
\providecommand \natexlab [1]{#1}%
\providecommand \enquote  [1]{``#1''}%
\providecommand \bibnamefont  [1]{#1}%
\providecommand \bibfnamefont [1]{#1}%
\providecommand \citenamefont [1]{#1}%
\providecommand \href@noop [0]{\@secondoftwo}%
\providecommand \href [0]{\begingroup \@sanitize@url \@href}%
\providecommand \@href[1]{\@@startlink{#1}\@@href}%
\providecommand \@@href[1]{\endgroup#1\@@endlink}%
\providecommand \@sanitize@url [0]{\catcode `\\12\catcode `\$12\catcode
  `\&12\catcode `\#12\catcode `\^12\catcode `\_12\catcode `\%12\relax}%
\providecommand \@@startlink[1]{}%
\providecommand \@@endlink[0]{}%
\providecommand \url  [0]{\begingroup\@sanitize@url \@url }%
\providecommand \@url [1]{\endgroup\@href {#1}{\urlprefix }}%
\providecommand \urlprefix  [0]{URL }%
\providecommand \Eprint [0]{\href }%
\providecommand \doibase [0]{http://dx.doi.org/}%
\providecommand \selectlanguage [0]{\@gobble}%
\providecommand \bibinfo  [0]{\@secondoftwo}%
\providecommand \bibfield  [0]{\@secondoftwo}%
\providecommand \translation [1]{[#1]}%
\providecommand \BibitemOpen [0]{}%
\providecommand \bibitemStop [0]{}%
\providecommand \bibitemNoStop [0]{.\EOS\space}%
\providecommand \EOS [0]{\spacefactor3000\relax}%
\providecommand \BibitemShut  [1]{\csname bibitem#1\endcsname}%
\let\auto@bib@innerbib\@empty
\bibitem [{\citenamefont {Geim}\ and\ \citenamefont
  {Novoselov}(2007)}]{graph1}%
  \BibitemOpen
  \bibfield  {author} {\bibinfo {author} {\bibfnamefont {A.~K.}\ \bibnamefont
  {Geim}}\ and\ \bibinfo {author} {\bibfnamefont {K.~S.}\ \bibnamefont
  {Novoselov}},\ }\href@noop {} {\bibfield  {journal} {\bibinfo  {journal}
  {Nat. Mater.}\ }\textbf {\bibinfo {volume} {6}},\ \bibinfo {pages} {183}
  (\bibinfo {year} {2007})}\BibitemShut {NoStop}%
\bibitem [{\citenamefont {Castro~Neto}\ \emph {et~al.}(2009)\citenamefont
  {Castro~Neto}, \citenamefont {Guinea}, \citenamefont {Peres}, \citenamefont
  {Novoselov},\ and\ \citenamefont {Geim}}]{graph2}%
  \BibitemOpen
  \bibfield  {author} {\bibinfo {author} {\bibfnamefont {A.~H.}\ \bibnamefont
  {Castro~Neto}}, \bibinfo {author} {\bibfnamefont {F.}~\bibnamefont {Guinea}},
  \bibinfo {author} {\bibfnamefont {N.~M.}\ \bibnamefont {Peres}}, \bibinfo
  {author} {\bibfnamefont {K.~S.}\ \bibnamefont {Novoselov}}, \ and\ \bibinfo
  {author} {\bibfnamefont {A.~K.}\ \bibnamefont {Geim}},\ }\href@noop {}
  {\bibfield  {journal} {\bibinfo  {journal} {Rev. Mod. Phys.}\ }\textbf
  {\bibinfo {volume} {81}},\ \bibinfo {pages} {109} (\bibinfo {year}
  {2009})}\BibitemShut {NoStop}%
\bibitem [{\citenamefont {Geim}(2009)}]{graph3}%
  \BibitemOpen
  \bibfield  {author} {\bibinfo {author} {\bibfnamefont {A.~K.}\ \bibnamefont
  {Geim}},\ }\href@noop {} {\bibfield  {journal} {\bibinfo  {journal}
  {Science}\ }\textbf {\bibinfo {volume} {324}},\ \bibinfo {pages} {1530}
  (\bibinfo {year} {2009})}\BibitemShut {NoStop}%
\bibitem [{\citenamefont {Sofo}\ \emph {et~al.}(2007)\citenamefont {Sofo},
  \citenamefont {Chaudhari},\ and\ \citenamefont {Barber}}]{sofo:153401}%
  \BibitemOpen
  \bibfield  {author} {\bibinfo {author} {\bibfnamefont {J.~O.}\ \bibnamefont
  {Sofo}}, \bibinfo {author} {\bibfnamefont {A.~S.}\ \bibnamefont {Chaudhari}},
  \ and\ \bibinfo {author} {\bibfnamefont {G.~D.}\ \bibnamefont {Barber}},\
  }\href@noop {} {\bibfield  {journal} {\bibinfo  {journal} {Phys Rev B}\
  }\textbf {\bibinfo {volume} {75}},\ \bibinfo {pages} {153401} (\bibinfo
  {year} {2007})}\BibitemShut {NoStop}%
\bibitem [{\citenamefont {Elias}\ \emph {et~al.}(2009)\citenamefont {Elias},
  \citenamefont {Nair}, \citenamefont {Mohiuddin}, \citenamefont {Morozov},
  \citenamefont {Blake}, \citenamefont {Halsall}, \citenamefont {Ferrari},
  \citenamefont {Boukhvalov}, \citenamefont {Katsnelson}, \citenamefont
  {Geim},\ and\ \citenamefont {Novoselov}}]{elias}%
  \BibitemOpen
  \bibfield  {author} {\bibinfo {author} {\bibfnamefont {D.~C.}\ \bibnamefont
  {Elias}}, \bibinfo {author} {\bibfnamefont {R.~R.}\ \bibnamefont {Nair}},
  \bibinfo {author} {\bibfnamefont {T.~M.~G.}\ \bibnamefont {Mohiuddin}},
  \bibinfo {author} {\bibfnamefont {S.~V.}\ \bibnamefont {Morozov}}, \bibinfo
  {author} {\bibfnamefont {P.}~\bibnamefont {Blake}}, \bibinfo {author}
  {\bibfnamefont {M.~P.}\ \bibnamefont {Halsall}}, \bibinfo {author}
  {\bibfnamefont {A.~C.}\ \bibnamefont {Ferrari}}, \bibinfo {author}
  {\bibfnamefont {D.~W.}\ \bibnamefont {Boukhvalov}}, \bibinfo {author}
  {\bibfnamefont {M.~I.}\ \bibnamefont {Katsnelson}}, \bibinfo {author}
  {\bibfnamefont {A.~K.}\ \bibnamefont {Geim}}, \ and\ \bibinfo {author}
  {\bibfnamefont {K.~S.}\ \bibnamefont {Novoselov}},\ }\href@noop {} {\bibfield
   {journal} {\bibinfo  {journal} {Science}\ }\textbf {\bibinfo {volume}
  {323}},\ \bibinfo {pages} {610} (\bibinfo {year} {2009})}\BibitemShut
  {NoStop}%
\bibitem [{\citenamefont {Son}\ \emph {et~al.}(2006{\natexlab{a}})\citenamefont
  {Son}, \citenamefont {Cohen},\ and\ \citenamefont {Louie}}]{loui}%
  \BibitemOpen
  \bibfield  {author} {\bibinfo {author} {\bibfnamefont {Y.-W.}\ \bibnamefont
  {Son}}, \bibinfo {author} {\bibfnamefont {M.~L.}\ \bibnamefont {Cohen}}, \
  and\ \bibinfo {author} {\bibfnamefont {S.~G.}\ \bibnamefont {Louie}},\
  }\href@noop {} {\bibfield  {journal} {\bibinfo  {journal} {Phys. Rev. Lett.}\
  }\textbf {\bibinfo {volume} {97}},\ \bibinfo {pages} {216803} (\bibinfo
  {year} {2006}{\natexlab{a}})}\BibitemShut {NoStop}%
\bibitem [{\citenamefont {Son}\ \emph {et~al.}(2006{\natexlab{b}})\citenamefont
  {Son}, \citenamefont {Cohen},\ and\ \citenamefont {Louie}}]{cohen}%
  \BibitemOpen
  \bibfield  {author} {\bibinfo {author} {\bibfnamefont {Y.-W.}\ \bibnamefont
  {Son}}, \bibinfo {author} {\bibfnamefont {M.~L.}\ \bibnamefont {Cohen}}, \
  and\ \bibinfo {author} {\bibfnamefont {S.~G.}\ \bibnamefont {Louie}},\
  }\href@noop {} {\bibfield  {journal} {\bibinfo  {journal} {Nature}\ }\textbf
  {\bibinfo {volume} {444}},\ \bibinfo {pages} {347} (\bibinfo {year}
  {2006}{\natexlab{b}})}\BibitemShut {NoStop}%
\bibitem [{\citenamefont {Han}\ \emph {et~al.}(2007)\citenamefont {Han},
  \citenamefont {\"Ozyilmaz}, \citenamefont {Zhang},\ and\ \citenamefont
  {Kim}}]{kim-prl}%
  \BibitemOpen
  \bibfield  {author} {\bibinfo {author} {\bibfnamefont {M.~Y.}\ \bibnamefont
  {Han}}, \bibinfo {author} {\bibfnamefont {B.}~\bibnamefont {\"Ozyilmaz}},
  \bibinfo {author} {\bibfnamefont {Y.}~\bibnamefont {Zhang}}, \ and\ \bibinfo
  {author} {\bibfnamefont {P.}~\bibnamefont {Kim}},\ }\href@noop {} {\bibfield
  {journal} {\bibinfo  {journal} {Phys. Rev. Lett.}\ }\textbf {\bibinfo
  {volume} {98}},\ \bibinfo {pages} {206805} (\bibinfo {year}
  {2007})}\BibitemShut {NoStop}%
\bibitem [{\citenamefont {Jiao}\ \emph {et~al.}(2009)\citenamefont {Jiao},
  \citenamefont {Zhang}, \citenamefont {Wang}, \citenamefont {Diankov},\ and\
  \citenamefont {Dai}}]{dai}%
  \BibitemOpen
  \bibfield  {author} {\bibinfo {author} {\bibfnamefont {L.}~\bibnamefont
  {Jiao}}, \bibinfo {author} {\bibfnamefont {L.}~\bibnamefont {Zhang}},
  \bibinfo {author} {\bibfnamefont {X.}~\bibnamefont {Wang}}, \bibinfo {author}
  {\bibfnamefont {G.}~\bibnamefont {Diankov}}, \ and\ \bibinfo {author}
  {\bibfnamefont {H.}~\bibnamefont {Dai}},\ }\href@noop {} {\bibfield
  {journal} {\bibinfo  {journal} {NATURE}\ }\textbf {\bibinfo {volume} {458}},\
  \bibinfo {pages} {877} (\bibinfo {year} {2009})}\BibitemShut {NoStop}%
\bibitem [{\citenamefont {Boukhvalov}\ \emph {et~al.}(2008)\citenamefont
  {Boukhvalov}, \citenamefont {Katsnelson},\ and\ \citenamefont
  {Lichtenstein}}]{boukhvalov}%
  \BibitemOpen
  \bibfield  {author} {\bibinfo {author} {\bibfnamefont {D.~W.}\ \bibnamefont
  {Boukhvalov}}, \bibinfo {author} {\bibfnamefont {M.~I.}\ \bibnamefont
  {Katsnelson}}, \ and\ \bibinfo {author} {\bibfnamefont {A.~I.}\ \bibnamefont
  {Lichtenstein}},\ }\href@noop {} {\bibfield  {journal} {\bibinfo  {journal}
  {Phys Rev B}\ }\textbf {\bibinfo {volume} {77}},\ \bibinfo {pages} {035427}
  (\bibinfo {year} {2008})}\BibitemShut {NoStop}%
\bibitem [{\citenamefont {Casolo}\ \emph {et~al.}(2009)\citenamefont {Casolo},
  \citenamefont {L{\o}vvik}, \citenamefont {Martinazzo},\ and\ \citenamefont
  {Tantardini}}]{casolo}%
  \BibitemOpen
  \bibfield  {author} {\bibinfo {author} {\bibfnamefont {S.}~\bibnamefont
  {Casolo}}, \bibinfo {author} {\bibfnamefont {O.~M.}\ \bibnamefont
  {L{\o}vvik}}, \bibinfo {author} {\bibfnamefont {R.}~\bibnamefont
  {Martinazzo}}, \ and\ \bibinfo {author} {\bibfnamefont {G.~F.}\ \bibnamefont
  {Tantardini}},\ }\href@noop {} {\bibfield  {journal} {\bibinfo  {journal} {J.
  Chem. Phys.}\ }\textbf {\bibinfo {volume} {130}},\ \bibinfo {pages} {054704}
  (\bibinfo {year} {2009})}\BibitemShut {NoStop}%
\bibitem [{\citenamefont {Leb{\`e}gue}\ \emph {et~al.}(2009)\citenamefont
  {Leb{\`e}gue}, \citenamefont {Klintenberg}, \citenamefont {Eriksson},\ and\
  \citenamefont {Katsnelson}}]{lebegue}%
  \BibitemOpen
  \bibfield  {author} {\bibinfo {author} {\bibfnamefont {S.}~\bibnamefont
  {Leb{\`e}gue}}, \bibinfo {author} {\bibfnamefont {M.}~\bibnamefont
  {Klintenberg}}, \bibinfo {author} {\bibfnamefont {O.}~\bibnamefont
  {Eriksson}}, \ and\ \bibinfo {author} {\bibfnamefont {M.~I.}\ \bibnamefont
  {Katsnelson}},\ }\href@noop {} {\bibfield  {journal} {\bibinfo  {journal}
  {Phys Rev B}\ }\textbf {\bibinfo {volume} {79}},\ \bibinfo {pages} {245117}
  (\bibinfo {year} {2009})}\BibitemShut {NoStop}%
\bibitem [{\citenamefont {Flores}\ \emph {et~al.}(2009)\citenamefont {Flores},
  \citenamefont {Autreto}, \citenamefont {Legoas},\ and\ \citenamefont
  {Galvao}}]{flores}%
  \BibitemOpen
  \bibfield  {author} {\bibinfo {author} {\bibfnamefont {M.~Z.~S.}\
  \bibnamefont {Flores}}, \bibinfo {author} {\bibfnamefont {P.~A.~S.}\
  \bibnamefont {Autreto}}, \bibinfo {author} {\bibfnamefont {S.~B.}\
  \bibnamefont {Legoas}}, \ and\ \bibinfo {author} {\bibfnamefont {D.~S.}\
  \bibnamefont {Galvao}},\ }\href@noop {} {\bibfield  {journal} {\bibinfo
  {journal} {Nanotechnology}\ }\textbf {\bibinfo {volume} {20}},\ \bibinfo
  {pages} {465704} (\bibinfo {year} {2009})}\BibitemShut {NoStop}%
\bibitem [{\citenamefont {Wu}\ \emph {et~al.}(2010)\citenamefont {Wu},
  \citenamefont {Wu}, \citenamefont {Gao},\ and\ \citenamefont {Zeng}}]{Wu}%
  \BibitemOpen
  \bibfield  {author} {\bibinfo {author} {\bibfnamefont {M.}~\bibnamefont
  {Wu}}, \bibinfo {author} {\bibfnamefont {X.}~\bibnamefont {Wu}}, \bibinfo
  {author} {\bibfnamefont {Y.}~\bibnamefont {Gao}}, \ and\ \bibinfo {author}
  {\bibfnamefont {X.~C.}\ \bibnamefont {Zeng}},\ }\href@noop {} {\bibfield
  {journal} {\bibinfo  {journal} {The Journal of Physical Chemistry C}\
  }\textbf {\bibinfo {volume} {114}},\ \bibinfo {pages} {139} (\bibinfo {year}
  {2010})}\BibitemShut {NoStop}%
\bibitem [{\citenamefont {\c{S}ahin}\ \emph {et~al.}(2009)\citenamefont
  {\c{S}ahin}, \citenamefont {Ataca},\ and\ \citenamefont
  {Ciraci}}]{sahin-apl}%
  \BibitemOpen
  \bibfield  {author} {\bibinfo {author} {\bibfnamefont {H.}~\bibnamefont
  {\c{S}ahin}}, \bibinfo {author} {\bibfnamefont {C.}~\bibnamefont {Ataca}}, \
  and\ \bibinfo {author} {\bibfnamefont {S.}~\bibnamefont {Ciraci}},\
  }\href@noop {} {\bibfield  {journal} {\bibinfo  {journal} {Applied Physics
  Letters}\ }\textbf {\bibinfo {volume} {95}},\ \bibinfo {pages} {222510}
  (\bibinfo {year} {2009})}\BibitemShut {NoStop}%
\bibitem [{\citenamefont {Robinson}\ \emph {et~al.}(2010)\citenamefont
  {Robinson}, \citenamefont {Burgess}, \citenamefont {Junkermeier},
  \citenamefont {Badescu}, \citenamefont {Reinecke}, \citenamefont {Perkins},
  \citenamefont {Zalalutdniov}, \citenamefont {Baldwin}, \citenamefont
  {Culbertson}, \citenamefont {Sheehan},\ and\ \citenamefont {Snow}}]{nrl-cf}%
  \BibitemOpen
  \bibfield  {author} {\bibinfo {author} {\bibfnamefont {J.~T.}\ \bibnamefont
  {Robinson}}, \bibinfo {author} {\bibfnamefont {J.~S.}\ \bibnamefont
  {Burgess}}, \bibinfo {author} {\bibfnamefont {C.~E.}\ \bibnamefont
  {Junkermeier}}, \bibinfo {author} {\bibfnamefont {S.~C.}\ \bibnamefont
  {Badescu}}, \bibinfo {author} {\bibfnamefont {T.~L.}\ \bibnamefont
  {Reinecke}}, \bibinfo {author} {\bibfnamefont {F.~K.}\ \bibnamefont
  {Perkins}}, \bibinfo {author} {\bibfnamefont {M.~K.}\ \bibnamefont
  {Zalalutdniov}}, \bibinfo {author} {\bibfnamefont {J.~W.}\ \bibnamefont
  {Baldwin}}, \bibinfo {author} {\bibfnamefont {J.~C.}\ \bibnamefont
  {Culbertson}}, \bibinfo {author} {\bibfnamefont {P.~E.}\ \bibnamefont
  {Sheehan}}, \ and\ \bibinfo {author} {\bibfnamefont {E.~S.}\ \bibnamefont
  {Snow}},\ }\href@noop {} {\bibfield  {journal} {\bibinfo  {journal} {Nano
  Letters}\ }\textbf {\bibinfo {volume} {10}},\ \bibinfo {pages} {3001}
  (\bibinfo {year} {2010})}\BibitemShut {NoStop}%
\bibitem [{\citenamefont {Leenaerts}\ \emph {et~al.}(2010)\citenamefont
  {Leenaerts}, \citenamefont {Peelaers}, \citenamefont {Hern\'andez-Nieves},
  \citenamefont {Partoens},\ and\ \citenamefont {Peeters}}]{peeters-cf}%
  \BibitemOpen
  \bibfield  {author} {\bibinfo {author} {\bibfnamefont {O.}~\bibnamefont
  {Leenaerts}}, \bibinfo {author} {\bibfnamefont {H.}~\bibnamefont {Peelaers}},
  \bibinfo {author} {\bibfnamefont {A.~D.}\ \bibnamefont {Hern\'andez-Nieves}},
  \bibinfo {author} {\bibfnamefont {B.}~\bibnamefont {Partoens}}, \ and\
  \bibinfo {author} {\bibfnamefont {F.~M.}\ \bibnamefont {Peeters}},\
  }\href@noop {} {\bibfield  {journal} {\bibinfo  {journal} {Phys. Rev. B}\
  }\textbf {\bibinfo {volume} {82}},\ \bibinfo {pages} {195436} (\bibinfo
  {year} {2010})}\BibitemShut {NoStop}%
\bibitem [{\citenamefont {Nair}\ \emph {et~al.}(2010)\citenamefont {Nair},
  \citenamefont {Ren}, \citenamefont {Jalil}, \citenamefont {Riaz},
  \citenamefont {Kravets}, \citenamefont {Britnell}, \citenamefont {Blake},
  \citenamefont {Schedin}, \citenamefont {Mayorov}, \citenamefont {Yuan},
  \citenamefont {Katsnelson}, \citenamefont {Cheng}, \citenamefont
  {Strupinski}, \citenamefont {Bulusheva}, \citenamefont {Okotrub},
  \citenamefont {Grigorieva}, \citenamefont {Grigorenko}, \citenamefont
  {Novoselov},\ and\ \citenamefont {Geim}}]{geim-small}%
  \BibitemOpen
  \bibfield  {author} {\bibinfo {author} {\bibfnamefont {R.~R.}\ \bibnamefont
  {Nair}}, \bibinfo {author} {\bibfnamefont {W.}~\bibnamefont {Ren}}, \bibinfo
  {author} {\bibfnamefont {R.}~\bibnamefont {Jalil}}, \bibinfo {author}
  {\bibfnamefont {I.}~\bibnamefont {Riaz}}, \bibinfo {author} {\bibfnamefont
  {V.~G.}\ \bibnamefont {Kravets}}, \bibinfo {author} {\bibfnamefont
  {L.}~\bibnamefont {Britnell}}, \bibinfo {author} {\bibfnamefont
  {P.}~\bibnamefont {Blake}}, \bibinfo {author} {\bibfnamefont
  {F.}~\bibnamefont {Schedin}}, \bibinfo {author} {\bibfnamefont {A.~S.}\
  \bibnamefont {Mayorov}}, \bibinfo {author} {\bibfnamefont {S.}~\bibnamefont
  {Yuan}}, \bibinfo {author} {\bibfnamefont {M.~I.}\ \bibnamefont
  {Katsnelson}}, \bibinfo {author} {\bibfnamefont {H.-M.}\ \bibnamefont
  {Cheng}}, \bibinfo {author} {\bibfnamefont {W.}~\bibnamefont {Strupinski}},
  \bibinfo {author} {\bibfnamefont {L.~G.}\ \bibnamefont {Bulusheva}}, \bibinfo
  {author} {\bibfnamefont {A.~V.}\ \bibnamefont {Okotrub}}, \bibinfo {author}
  {\bibfnamefont {I.~V.}\ \bibnamefont {Grigorieva}}, \bibinfo {author}
  {\bibfnamefont {A.~N.}\ \bibnamefont {Grigorenko}}, \bibinfo {author}
  {\bibfnamefont {K.~S.}\ \bibnamefont {Novoselov}}, \ and\ \bibinfo {author}
  {\bibfnamefont {A.~K.}\ \bibnamefont {Geim}},\ }\href@noop {} {\bibfield
  {journal} {\bibinfo  {journal} {Small}\ }\textbf {\bibinfo {volume} {6}},\
  \bibinfo {pages} {2877} (\bibinfo {year} {2010})}\BibitemShut {NoStop}%
\bibitem [{\citenamefont {Samarakoon}\ \emph {et~al.}(2011)\citenamefont
  {Samarakoon}, \citenamefont {Chen}, \citenamefont {Nicolas},\ and\
  \citenamefont {Wang}}]{wang-small}%
  \BibitemOpen
  \bibfield  {author} {\bibinfo {author} {\bibfnamefont {D.~K.}\ \bibnamefont
  {Samarakoon}}, \bibinfo {author} {\bibfnamefont {Z.}~\bibnamefont {Chen}},
  \bibinfo {author} {\bibfnamefont {C.}~\bibnamefont {Nicolas}}, \ and\
  \bibinfo {author} {\bibfnamefont {X.-Q.}\ \bibnamefont {Wang}},\ }\href@noop
  {} {\bibfield  {journal} {\bibinfo  {journal} {Small}\ }\textbf {\bibinfo
  {volume} {7}},\ \bibinfo {pages} {965} (\bibinfo {year} {2011})}\BibitemShut
  {NoStop}%
\bibitem [{\citenamefont {Cheng}\ \emph {et~al.}(2010)\citenamefont {Cheng},
  \citenamefont {Zou}, \citenamefont {Okino}, \citenamefont {Gutierrez},
  \citenamefont {Gupta}, \citenamefont {Shen}, \citenamefont {Eklund},
  \citenamefont {Sofo},\ and\ \citenamefont {Zhu}}]{zhu-prb}%
  \BibitemOpen
  \bibfield  {author} {\bibinfo {author} {\bibfnamefont {S.-H.}\ \bibnamefont
  {Cheng}}, \bibinfo {author} {\bibfnamefont {K.}~\bibnamefont {Zou}}, \bibinfo
  {author} {\bibfnamefont {F.}~\bibnamefont {Okino}}, \bibinfo {author}
  {\bibfnamefont {H.~R.}\ \bibnamefont {Gutierrez}}, \bibinfo {author}
  {\bibfnamefont {A.}~\bibnamefont {Gupta}}, \bibinfo {author} {\bibfnamefont
  {N.}~\bibnamefont {Shen}}, \bibinfo {author} {\bibfnamefont {P.~C.}\
  \bibnamefont {Eklund}}, \bibinfo {author} {\bibfnamefont {J.~O.}\
  \bibnamefont {Sofo}}, \ and\ \bibinfo {author} {\bibfnamefont
  {J.}~\bibnamefont {Zhu}},\ }\href@noop {} {\bibfield  {journal} {\bibinfo
  {journal} {Phys. Rev. B}\ }\textbf {\bibinfo {volume} {81}},\ \bibinfo
  {pages} {205435} (\bibinfo {year} {2010})}\BibitemShut {NoStop}%
\bibitem [{\citenamefont {\ifmmode~\mbox{\c{S}}\else \c{S}\fi{}ahin}\ \emph
  {et~al.}(2011)\citenamefont {\ifmmode~\mbox{\c{S}}\else \c{S}\fi{}ahin},
  \citenamefont {Topsakal},\ and\ \citenamefont {Ciraci}}]{ciraci-cf}%
  \BibitemOpen
  \bibfield  {author} {\bibinfo {author} {\bibfnamefont {H.}~\bibnamefont
  {\ifmmode~\mbox{\c{S}}\else \c{S}\fi{}ahin}}, \bibinfo {author}
  {\bibfnamefont {M.}~\bibnamefont {Topsakal}}, \ and\ \bibinfo {author}
  {\bibfnamefont {S.}~\bibnamefont {Ciraci}},\ }\href@noop {} {\bibfield
  {journal} {\bibinfo  {journal} {Phys. Rev. B}\ }\textbf {\bibinfo {volume}
  {83}},\ \bibinfo {pages} {115432} (\bibinfo {year} {2011})}\BibitemShut
  {NoStop}%
\bibitem [{\citenamefont {Chandrachud}\ \emph {et~al.}(2010)\citenamefont
  {Chandrachud}, \citenamefont {Pujari}, \citenamefont {Haldar}, \citenamefont
  {Sanyal},\ and\ \citenamefont {Kanhere}}]{prachi}%
  \BibitemOpen
  \bibfield  {author} {\bibinfo {author} {\bibfnamefont {P.}~\bibnamefont
  {Chandrachud}}, \bibinfo {author} {\bibfnamefont {B.~S.}\ \bibnamefont
  {Pujari}}, \bibinfo {author} {\bibfnamefont {S.}~\bibnamefont {Haldar}},
  \bibinfo {author} {\bibfnamefont {B.}~\bibnamefont {Sanyal}}, \ and\ \bibinfo
  {author} {\bibfnamefont {D.~G.}\ \bibnamefont {Kanhere}},\ }\href@noop {}
  {\bibfield  {journal} {\bibinfo  {journal} {Journal of Physics: Condensed
  Matter}\ }\textbf {\bibinfo {volume} {22}},\ \bibinfo {pages} {465502}
  (\bibinfo {year} {2010})}\BibitemShut {NoStop}%
\bibitem [{\citenamefont {Singh}\ and\ \citenamefont
  {Yakobson}(2009)}]{aksingh}%
  \BibitemOpen
  \bibfield  {author} {\bibinfo {author} {\bibfnamefont {A.~K.}\ \bibnamefont
  {Singh}}\ and\ \bibinfo {author} {\bibfnamefont {B.~I.}\ \bibnamefont
  {Yakobson}},\ }\href@noop {} {\bibfield  {journal} {\bibinfo  {journal} {Nano
  Letters}\ }\textbf {\bibinfo {volume} {9}},\ \bibinfo {pages} {1540}
  (\bibinfo {year} {2009})}\BibitemShut {NoStop}%
\bibitem [{\citenamefont {Sessi}\ \emph {et~al.}(2009)\citenamefont {Sessi},
  \citenamefont {Guest}, \citenamefont {Bode},\ and\ \citenamefont
  {Guisinger}}]{sessi}%
  \BibitemOpen
  \bibfield  {author} {\bibinfo {author} {\bibfnamefont {P.}~\bibnamefont
  {Sessi}}, \bibinfo {author} {\bibfnamefont {J.~R.}\ \bibnamefont {Guest}},
  \bibinfo {author} {\bibfnamefont {M.}~\bibnamefont {Bode}}, \ and\ \bibinfo
  {author} {\bibfnamefont {N.~P.}\ \bibnamefont {Guisinger}},\ }\href@noop {}
  {\bibfield  {journal} {\bibinfo  {journal} {Nano Letters}\ }\textbf {\bibinfo
  {volume} {9}},\ \bibinfo {pages} {4343} (\bibinfo {year} {2009})}\BibitemShut
  {NoStop}%
\bibitem [{\citenamefont {Lu}\ and\ \citenamefont {Feng}(2009)}]{ylu}%
  \BibitemOpen
  \bibfield  {author} {\bibinfo {author} {\bibfnamefont {Y.~H.}\ \bibnamefont
  {Lu}}\ and\ \bibinfo {author} {\bibfnamefont {Y.~P.}\ \bibnamefont {Feng}},\
  }\href@noop {} {\bibfield  {journal} {\bibinfo  {journal} {The Journal of
  Physical Chemistry C}\ }\textbf {\bibinfo {volume} {113}},\ \bibinfo {pages}
  {20841} (\bibinfo {year} {2009})}\BibitemShut {NoStop}%
\bibitem [{\citenamefont {Balog}\ \emph {et~al.}(2010)\citenamefont {Balog},
  \citenamefont {J\o~rgensen}, \citenamefont {Nilsson}, \citenamefont
  {Andersen}, \citenamefont {Rienks}, \citenamefont {Bianchi}, \citenamefont
  {Fanetti}, \citenamefont {Laegsgaard}, \citenamefont {Baraldi}, \citenamefont
  {Lizzit}, \citenamefont {Sljivancanin}, \citenamefont {Besenbacher},
  \citenamefont {Hammer}, \citenamefont {Pedersen}, \citenamefont {Hofmann},\
  and\ \citenamefont {Hornekaer}}]{balog}%
  \BibitemOpen
  \bibfield  {author} {\bibinfo {author} {\bibfnamefont {R.}~\bibnamefont
  {Balog}}, \bibinfo {author} {\bibfnamefont {B.}~\bibnamefont {J\o~rgensen}},
  \bibinfo {author} {\bibfnamefont {L.}~\bibnamefont {Nilsson}}, \bibinfo
  {author} {\bibfnamefont {M.}~\bibnamefont {Andersen}}, \bibinfo {author}
  {\bibfnamefont {E.}~\bibnamefont {Rienks}}, \bibinfo {author} {\bibfnamefont
  {M.}~\bibnamefont {Bianchi}}, \bibinfo {author} {\bibfnamefont
  {M.}~\bibnamefont {Fanetti}}, \bibinfo {author} {\bibfnamefont
  {E.}~\bibnamefont {Laegsgaard}}, \bibinfo {author} {\bibfnamefont
  {A.}~\bibnamefont {Baraldi}}, \bibinfo {author} {\bibfnamefont
  {S.}~\bibnamefont {Lizzit}}, \bibinfo {author} {\bibfnamefont
  {Z.}~\bibnamefont {Sljivancanin}}, \bibinfo {author} {\bibfnamefont
  {F.}~\bibnamefont {Besenbacher}}, \bibinfo {author} {\bibfnamefont {B.~r.}\
  \bibnamefont {Hammer}}, \bibinfo {author} {\bibfnamefont {T.~G.}\
  \bibnamefont {Pedersen}}, \bibinfo {author} {\bibfnamefont {P.}~\bibnamefont
  {Hofmann}}, \ and\ \bibinfo {author} {\bibfnamefont {L.}~\bibnamefont
  {Hornekaer}},\ }\href@noop {} {\bibfield  {journal} {\bibinfo  {journal}
  {Nature materials}\ }\textbf {\bibinfo {volume} {9}},\ \bibinfo {pages} {315}
  (\bibinfo {year} {2010})}\BibitemShut {NoStop}%
\bibitem [{\citenamefont {Ao}\ \emph {et~al.}(2010)\citenamefont {Ao},
  \citenamefont {Hern\'andez-Nieves}, \citenamefont {Peeters},\ and\
  \citenamefont {Li}}]{zm-ao}%
  \BibitemOpen
  \bibfield  {author} {\bibinfo {author} {\bibfnamefont {Z.~M.}\ \bibnamefont
  {Ao}}, \bibinfo {author} {\bibfnamefont {A.~D.}\ \bibnamefont
  {Hern\'andez-Nieves}}, \bibinfo {author} {\bibfnamefont {F.~M.}\ \bibnamefont
  {Peeters}}, \ and\ \bibinfo {author} {\bibfnamefont {S.}~\bibnamefont {Li}},\
  }\href@noop {} {\bibfield  {journal} {\bibinfo  {journal} {Appl. Phys.
  Lett.}\ }\textbf {\bibinfo {volume} {97}},\ \bibinfo {pages} {233109}
  (\bibinfo {year} {2010})}\BibitemShut {NoStop}%
\bibitem [{\citenamefont {Li}\ \emph {et~al.}(2009)\citenamefont {Li},
  \citenamefont {Zhou}, \citenamefont {Shen},\ and\ \citenamefont
  {Chen}}]{yli}%
  \BibitemOpen
  \bibfield  {author} {\bibinfo {author} {\bibfnamefont {Y.}~\bibnamefont
  {Li}}, \bibinfo {author} {\bibfnamefont {Z.}~\bibnamefont {Zhou}}, \bibinfo
  {author} {\bibfnamefont {P.}~\bibnamefont {Shen}}, \ and\ \bibinfo {author}
  {\bibfnamefont {Z.}~\bibnamefont {Chen}},\ }\href@noop {} {\bibfield
  {journal} {\bibinfo  {journal} {The Journal of Physical Chemistry C}\
  }\textbf {\bibinfo {volume} {113}},\ \bibinfo {pages} {15043} (\bibinfo
  {year} {2009})}\BibitemShut {NoStop}%
\bibitem [{\citenamefont {Rajabpour}\ \emph {et~al.}(2011)\citenamefont
  {Rajabpour}, \citenamefont {Allaei},\ and\ \citenamefont
  {Kowsary}}]{kowsary}%
  \BibitemOpen
  \bibfield  {author} {\bibinfo {author} {\bibfnamefont {A.}~\bibnamefont
  {Rajabpour}}, \bibinfo {author} {\bibfnamefont {S.~M.~V.}\ \bibnamefont
  {Allaei}}, \ and\ \bibinfo {author} {\bibfnamefont {F.}~\bibnamefont
  {Kowsary}},\ }\href@noop {} {\bibfield  {journal} {\bibinfo  {journal} {Appl.
  Phys. Lett.}\ }\textbf {\bibinfo {volume} {99}},\ \bibinfo {pages} {051917}
  (\bibinfo {year} {2011})}\BibitemShut {NoStop}%
\bibitem [{\citenamefont {Ij\"as}\ \emph {et~al.}(2011)\citenamefont {Ij\"as},
  \citenamefont {Havu}, \citenamefont {Harju},\ and\ \citenamefont
  {Pasanen}}]{ijas}%
  \BibitemOpen
  \bibfield  {author} {\bibinfo {author} {\bibfnamefont {M.}~\bibnamefont
  {Ij\"as}}, \bibinfo {author} {\bibfnamefont {P.}~\bibnamefont {Havu}},
  \bibinfo {author} {\bibfnamefont {A.}~\bibnamefont {Harju}}, \ and\ \bibinfo
  {author} {\bibfnamefont {P.}~\bibnamefont {Pasanen}},\ }\href@noop {}
  {\bibfield  {journal} {\bibinfo  {journal} {Phys. Rev. B}\ }\textbf {\bibinfo
  {volume} {84}},\ \bibinfo {pages} {041403} (\bibinfo {year}
  {2011})}\BibitemShut {NoStop}%
\bibitem [{\citenamefont {Yazyev}(2010)}]{yazev-rev}%
  \BibitemOpen
  \bibfield  {author} {\bibinfo {author} {\bibfnamefont {O.~V.}\ \bibnamefont
  {Yazyev}},\ }\href {http://stacks.iop.org/0034-4885/73/i=5/a=056501}
  {\bibfield  {journal} {\bibinfo  {journal} {Reports on Progress in Physics}\
  }\textbf {\bibinfo {volume} {73}},\ \bibinfo {pages} {056501} (\bibinfo
  {year} {2010})}\BibitemShut {NoStop}%
\bibitem [{\citenamefont {Nakada}\ \emph {et~al.}(1996)\citenamefont {Nakada},
  \citenamefont {Fujita}, \citenamefont {Dresselhaus},\ and\ \citenamefont
  {Dresselhaus}}]{nakada}%
  \BibitemOpen
  \bibfield  {author} {\bibinfo {author} {\bibfnamefont {K.}~\bibnamefont
  {Nakada}}, \bibinfo {author} {\bibfnamefont {M.}~\bibnamefont {Fujita}},
  \bibinfo {author} {\bibfnamefont {G.}~\bibnamefont {Dresselhaus}}, \ and\
  \bibinfo {author} {\bibfnamefont {M.~S.}\ \bibnamefont {Dresselhaus}},\
  }\href {\doibase 10.1103/PhysRevB.54.17954} {\bibfield  {journal} {\bibinfo
  {journal} {Phys. Rev. B}\ }\textbf {\bibinfo {volume} {54}},\ \bibinfo
  {pages} {17954} (\bibinfo {year} {1996})}\BibitemShut {NoStop}%
\bibitem [{\citenamefont {Jia}\ \emph {et~al.}(2009)\citenamefont {Jia},
  \citenamefont {Hofmann}, \citenamefont {Meunier}, \citenamefont {Sumpter},
  \citenamefont {Campos-Delgado}, \citenamefont {Romo-Herrera}, \citenamefont
  {Son}, \citenamefont {Hsieh}, \citenamefont {Reina}, \citenamefont {Kong},
  \citenamefont {Terrones},\ and\ \citenamefont {Dresselhaus}}]{hofman}%
  \BibitemOpen
  \bibfield  {author} {\bibinfo {author} {\bibfnamefont {X.}~\bibnamefont
  {Jia}}, \bibinfo {author} {\bibfnamefont {M.}~\bibnamefont {Hofmann}},
  \bibinfo {author} {\bibfnamefont {V.}~\bibnamefont {Meunier}}, \bibinfo
  {author} {\bibfnamefont {B.~G.}\ \bibnamefont {Sumpter}}, \bibinfo {author}
  {\bibfnamefont {J.}~\bibnamefont {Campos-Delgado}}, \bibinfo {author}
  {\bibfnamefont {J.~M.}\ \bibnamefont {Romo-Herrera}}, \bibinfo {author}
  {\bibfnamefont {H.}~\bibnamefont {Son}}, \bibinfo {author} {\bibfnamefont
  {Y.-P.}\ \bibnamefont {Hsieh}}, \bibinfo {author} {\bibfnamefont
  {A.}~\bibnamefont {Reina}}, \bibinfo {author} {\bibfnamefont
  {J.}~\bibnamefont {Kong}}, \bibinfo {author} {\bibfnamefont {M.}~\bibnamefont
  {Terrones}}, \ and\ \bibinfo {author} {\bibfnamefont {M.~S.}\ \bibnamefont
  {Dresselhaus}},\ }\href@noop {} {\bibfield  {journal} {\bibinfo  {journal}
  {Science}\ }\textbf {\bibinfo {volume} {323}},\ \bibinfo {pages} {1701}
  (\bibinfo {year} {2009})}\BibitemShut {NoStop}%
\bibitem [{\citenamefont {Girit}\ \emph {et~al.}(2009)\citenamefont {Girit},
  \citenamefont {Meyer}, \citenamefont {Erni}, \citenamefont {Rossell},
  \citenamefont {Kisielowski}, \citenamefont {Yang}, \citenamefont {Park},
  \citenamefont {Crommie}, \citenamefont {Cohen}, \citenamefont {Louie},\ and\
  \citenamefont {Zettl}}]{girit}%
  \BibitemOpen
  \bibfield  {author} {\bibinfo {author} {\bibfnamefont {c.~O.}\ \bibnamefont
  {Girit}}, \bibinfo {author} {\bibfnamefont {J.~C.}\ \bibnamefont {Meyer}},
  \bibinfo {author} {\bibfnamefont {R.}~\bibnamefont {Erni}}, \bibinfo {author}
  {\bibfnamefont {M.~D.}\ \bibnamefont {Rossell}}, \bibinfo {author}
  {\bibfnamefont {C.}~\bibnamefont {Kisielowski}}, \bibinfo {author}
  {\bibfnamefont {L.}~\bibnamefont {Yang}}, \bibinfo {author} {\bibfnamefont
  {C.-H.}\ \bibnamefont {Park}}, \bibinfo {author} {\bibfnamefont {M.~F.}\
  \bibnamefont {Crommie}}, \bibinfo {author} {\bibfnamefont {M.~L.}\
  \bibnamefont {Cohen}}, \bibinfo {author} {\bibfnamefont {S.~G.}\ \bibnamefont
  {Louie}}, \ and\ \bibinfo {author} {\bibfnamefont {A.}~\bibnamefont
  {Zettl}},\ }\href@noop {} {\bibfield  {journal} {\bibinfo  {journal}
  {Science}\ }\textbf {\bibinfo {volume} {323}},\ \bibinfo {pages} {1705}
  (\bibinfo {year} {2009})}\BibitemShut {NoStop}%
\bibitem [{\citenamefont {Koskinen}\ \emph {et~al.}(2009)\citenamefont
  {Koskinen}, \citenamefont {Malola},\ and\ \citenamefont
  {H\"akkinen}}]{koski-prb}%
  \BibitemOpen
  \bibfield  {author} {\bibinfo {author} {\bibfnamefont {P.}~\bibnamefont
  {Koskinen}}, \bibinfo {author} {\bibfnamefont {S.}~\bibnamefont {Malola}}, \
  and\ \bibinfo {author} {\bibfnamefont {H.}~\bibnamefont {H\"akkinen}},\
  }\href@noop {} {\bibfield  {journal} {\bibinfo  {journal} {Phys. Rev. B}\
  }\textbf {\bibinfo {volume} {80}},\ \bibinfo {pages} {073401} (\bibinfo
  {year} {2009})}\BibitemShut {NoStop}%
\bibitem [{\citenamefont {Koskinen}\ \emph {et~al.}(2008)\citenamefont
  {Koskinen}, \citenamefont {Malola},\ and\ \citenamefont
  {H\"akkinen}}]{koski-prl}%
  \BibitemOpen
  \bibfield  {author} {\bibinfo {author} {\bibfnamefont {P.}~\bibnamefont
  {Koskinen}}, \bibinfo {author} {\bibfnamefont {S.}~\bibnamefont {Malola}}, \
  and\ \bibinfo {author} {\bibfnamefont {H.}~\bibnamefont {H\"akkinen}},\
  }\href@noop {} {\bibfield  {journal} {\bibinfo  {journal} {Phys. Rev. Lett.}\
  }\textbf {\bibinfo {volume} {101}},\ \bibinfo {pages} {115502} (\bibinfo
  {year} {2008})}\BibitemShut {NoStop}%
\bibitem [{\citenamefont {Fern\'andez-Rossier}(2008)}]{mag2}%
  \BibitemOpen
  \bibfield  {author} {\bibinfo {author} {\bibfnamefont {J.}~\bibnamefont
  {Fern\'andez-Rossier}},\ }\href {\doibase 10.1103/PhysRevB.77.075430}
  {\bibfield  {journal} {\bibinfo  {journal} {Phys. Rev. B}\ }\textbf {\bibinfo
  {volume} {77}},\ \bibinfo {pages} {075430} (\bibinfo {year}
  {2008})}\BibitemShut {NoStop}%
\bibitem [{\citenamefont {Bhandary}\ \emph {et~al.}(2010)\citenamefont
  {Bhandary}, \citenamefont {Eriksson}, \citenamefont {Sanyal},\ and\
  \citenamefont {Katsnelson}}]{bhandary}%
  \BibitemOpen
  \bibfield  {author} {\bibinfo {author} {\bibfnamefont {S.}~\bibnamefont
  {Bhandary}}, \bibinfo {author} {\bibfnamefont {O.}~\bibnamefont {Eriksson}},
  \bibinfo {author} {\bibfnamefont {B.}~\bibnamefont {Sanyal}}, \ and\ \bibinfo
  {author} {\bibfnamefont {M.~I.}\ \bibnamefont {Katsnelson}},\ }\href@noop {}
  {\bibfield  {journal} {\bibinfo  {journal} {Phys. Rev. B}\ }\textbf {\bibinfo
  {volume} {82}},\ \bibinfo {pages} {165405} (\bibinfo {year}
  {2010})}\BibitemShut {NoStop}%
\bibitem [{\citenamefont {Kunstmann}\ \emph {et~al.}(2011)\citenamefont
  {Kunstmann}, \citenamefont {\"Ozdo\ifmmode~\breve{g}\else \u{g}\fi{}an},
  \citenamefont {Quandt},\ and\ \citenamefont {Fehske}}]{jens}%
  \BibitemOpen
  \bibfield  {author} {\bibinfo {author} {\bibfnamefont {J.}~\bibnamefont
  {Kunstmann}}, \bibinfo {author} {\bibfnamefont {C.}~\bibnamefont
  {\"Ozdo\ifmmode~\breve{g}\else \u{g}\fi{}an}}, \bibinfo {author}
  {\bibfnamefont {A.}~\bibnamefont {Quandt}}, \ and\ \bibinfo {author}
  {\bibfnamefont {H.}~\bibnamefont {Fehske}},\ }\href {\doibase
  10.1103/PhysRevB.83.045414} {\bibfield  {journal} {\bibinfo  {journal} {Phys.
  Rev. B}\ }\textbf {\bibinfo {volume} {83}},\ \bibinfo {pages} {045414}
  (\bibinfo {year} {2011})}\BibitemShut {NoStop}%
\bibitem [{\citenamefont {Wang}\ \emph {et~al.}(2010)\citenamefont {Wang},
  \citenamefont {Cao},\ and\ \citenamefont {Cheng}}]{haiping1}%
  \BibitemOpen
  \bibfield  {author} {\bibinfo {author} {\bibfnamefont {Y.}~\bibnamefont
  {Wang}}, \bibinfo {author} {\bibfnamefont {C.}~\bibnamefont {Cao}}, \ and\
  \bibinfo {author} {\bibfnamefont {H.-P.}\ \bibnamefont {Cheng}},\ }\href
  {\doibase 10.1103/PhysRevB.82.205429} {\bibfield  {journal} {\bibinfo
  {journal} {Phys. Rev. B}\ }\textbf {\bibinfo {volume} {82}},\ \bibinfo
  {pages} {205429} (\bibinfo {year} {2010})}\BibitemShut {NoStop}%
\bibitem [{\citenamefont {Wang}\ and\ \citenamefont {Cheng}(2011)}]{haiping2}%
  \BibitemOpen
  \bibfield  {author} {\bibinfo {author} {\bibfnamefont {Y.}~\bibnamefont
  {Wang}}\ and\ \bibinfo {author} {\bibfnamefont {H.-P.}\ \bibnamefont
  {Cheng}},\ }\href {\doibase 10.1103/PhysRevB.83.113402} {\bibfield  {journal}
  {\bibinfo  {journal} {Phys. Rev. B}\ }\textbf {\bibinfo {volume} {83}},\
  \bibinfo {pages} {113402} (\bibinfo {year} {2011})}\BibitemShut {NoStop}%
\bibitem [{\citenamefont {Power}\ \emph {et~al.}(2011)\citenamefont {Power},
  \citenamefont {de~Menezes}, \citenamefont {Fagan},\ and\ \citenamefont
  {Ferreira}}]{power}%
  \BibitemOpen
  \bibfield  {author} {\bibinfo {author} {\bibfnamefont {S.~R.}\ \bibnamefont
  {Power}}, \bibinfo {author} {\bibfnamefont {V.~M.}\ \bibnamefont
  {de~Menezes}}, \bibinfo {author} {\bibfnamefont {S.~B.}\ \bibnamefont
  {Fagan}}, \ and\ \bibinfo {author} {\bibfnamefont {M.~S.}\ \bibnamefont
  {Ferreira}},\ }\href {\doibase 10.1103/PhysRevB.84.195431} {\bibfield
  {journal} {\bibinfo  {journal} {Phys. Rev. B}\ }\textbf {\bibinfo {volume}
  {84}},\ \bibinfo {pages} {195431} (\bibinfo {year} {2011})}\BibitemShut
  {NoStop}%
\bibitem [{\citenamefont {Schmidt}\ and\ \citenamefont {Loss}(2010)}]{loss}%
  \BibitemOpen
  \bibfield  {author} {\bibinfo {author} {\bibfnamefont {M.~J.}\ \bibnamefont
  {Schmidt}}\ and\ \bibinfo {author} {\bibfnamefont {D.}~\bibnamefont {Loss}},\
  }\href@noop {} {\bibfield  {journal} {\bibinfo  {journal} {Phys. Rev. B}\
  }\textbf {\bibinfo {volume} {81}},\ \bibinfo {pages} {165439} (\bibinfo
  {year} {2010})}\BibitemShut {NoStop}%
\bibitem [{\citenamefont {Hern\'andez-Nieves}\ \emph
  {et~al.}(2010)\citenamefont {Hern\'andez-Nieves}, \citenamefont {Partoens},\
  and\ \citenamefont {Peeters}}]{nieves}%
  \BibitemOpen
  \bibfield  {author} {\bibinfo {author} {\bibfnamefont {A.~D.}\ \bibnamefont
  {Hern\'andez-Nieves}}, \bibinfo {author} {\bibfnamefont {B.}~\bibnamefont
  {Partoens}}, \ and\ \bibinfo {author} {\bibfnamefont {F.~M.}\ \bibnamefont
  {Peeters}},\ }\href@noop {} {\bibfield  {journal} {\bibinfo  {journal} {Phys.
  Rev. B}\ }\textbf {\bibinfo {volume} {82}},\ \bibinfo {pages} {165412}
  (\bibinfo {year} {2010})}\BibitemShut {NoStop}%
\bibitem [{\citenamefont {Bhandary}\ \emph {et~al.}(2011)\citenamefont
  {Bhandary}, \citenamefont {Ghosh}, \citenamefont {Herper}, \citenamefont
  {Wende}, \citenamefont {Eriksson},\ and\ \citenamefont {Sanyal}}]{fep}%
  \BibitemOpen
  \bibfield  {author} {\bibinfo {author} {\bibfnamefont {S.}~\bibnamefont
  {Bhandary}}, \bibinfo {author} {\bibfnamefont {S.}~\bibnamefont {Ghosh}},
  \bibinfo {author} {\bibfnamefont {H.}~\bibnamefont {Herper}}, \bibinfo
  {author} {\bibfnamefont {H.}~\bibnamefont {Wende}}, \bibinfo {author}
  {\bibfnamefont {O.}~\bibnamefont {Eriksson}}, \ and\ \bibinfo {author}
  {\bibfnamefont {B.}~\bibnamefont {Sanyal}},\ }\href@noop {} {\bibfield
  {journal} {\bibinfo  {journal} {Phys.\ Rev.\ Lett.}\ }\textbf {\bibinfo
  {volume} {107}},\ \bibinfo {pages} {257202} (\bibinfo {year}
  {2011})}\BibitemShut {NoStop}%
\bibitem [{\citenamefont {G.Kresse}\ and\ \citenamefont
  {Furthmuller}(1996{\natexlab{a}})}]{vasp}%
  \BibitemOpen
  \bibfield  {author} {\bibinfo {author} {\bibnamefont {G.Kresse}}\ and\
  \bibinfo {author} {\bibfnamefont {J.}~\bibnamefont {Furthmuller}},\
  }\href@noop {} {\bibfield  {journal} {\bibinfo  {journal} {Phys.\ Rev.\ B}\
  }\textbf {\bibinfo {volume} {54}},\ \bibinfo {pages} {11169} (\bibinfo {year}
  {1996}{\natexlab{a}})}\BibitemShut {NoStop}%
\bibitem [{\citenamefont {G.Kresse}\ and\ \citenamefont
  {Furthmuller}(1996{\natexlab{b}})}]{vasp1}%
  \BibitemOpen
  \bibfield  {author} {\bibinfo {author} {\bibnamefont {G.Kresse}}\ and\
  \bibinfo {author} {\bibfnamefont {J.}~\bibnamefont {Furthmuller}},\
  }\href@noop {} {\bibfield  {journal} {\bibinfo  {journal} {Comput. Mater.
  Sci.}\ }\textbf {\bibinfo {volume} {6}},\ \bibinfo {pages} {15} (\bibinfo
  {year} {1996}{\natexlab{b}})}\BibitemShut {NoStop}%
\bibitem [{\citenamefont {Bl\"ochl}(1994)}]{Blo}%
  \BibitemOpen
  \bibfield  {author} {\bibinfo {author} {\bibfnamefont {P.~E.}\ \bibnamefont
  {Bl\"ochl}},\ }\href@noop {} {\bibfield  {journal} {\bibinfo  {journal}
  {Phys.\ Rev.\ B}\ }\textbf {\bibinfo {volume} {50}},\ \bibinfo {pages}
  {17953} (\bibinfo {year} {1994})}\BibitemShut {NoStop}%
\bibitem [{\citenamefont {Kresse}\ and\ \citenamefont {Joubert}(1999)}]{Blo1}%
  \BibitemOpen
  \bibfield  {author} {\bibinfo {author} {\bibfnamefont {G.}~\bibnamefont
  {Kresse}}\ and\ \bibinfo {author} {\bibfnamefont {J.}~\bibnamefont
  {Joubert}},\ }\href@noop {} {\bibfield  {journal} {\bibinfo  {journal}
  {Phys.\ Rev.\ B}\ }\textbf {\bibinfo {volume} {59}},\ \bibinfo {pages} {1758}
  (\bibinfo {year} {1999})}\BibitemShut {NoStop}%
\bibitem [{\citenamefont {Perdew}\ \emph {et~al.}(1996)\citenamefont {Perdew},
  \citenamefont {Burke},\ and\ \citenamefont {Ernzerhof}}]{pbe}%
  \BibitemOpen
  \bibfield  {author} {\bibinfo {author} {\bibfnamefont {J.~P.}\ \bibnamefont
  {Perdew}}, \bibinfo {author} {\bibfnamefont {K.}~\bibnamefont {Burke}}, \
  and\ \bibinfo {author} {\bibfnamefont {M.}~\bibnamefont {Ernzerhof}},\
  }\href@noop {} {\bibfield  {journal} {\bibinfo  {journal} {Phys.\ Rev.\
  Lett.}\ }\textbf {\bibinfo {volume} {77}},\ \bibinfo {pages} {3865} (\bibinfo
  {year} {1996})}\BibitemShut {NoStop}%
\bibitem [{\citenamefont {Perdew}\ \emph {et~al.}(1997)\citenamefont {Perdew},
  \citenamefont {Burke},\ and\ \citenamefont {Ernzerhof}}]{pbe2}%
  \BibitemOpen
  \bibfield  {author} {\bibinfo {author} {\bibfnamefont {J.~P.}\ \bibnamefont
  {Perdew}}, \bibinfo {author} {\bibfnamefont {K.}~\bibnamefont {Burke}}, \
  and\ \bibinfo {author} {\bibfnamefont {M.}~\bibnamefont {Ernzerhof}},\ }\href
  {\doibase 10.1103/PhysRevLett.78.1396} {\bibfield  {journal} {\bibinfo
  {journal} {Phys. Rev. Lett.}\ }\textbf {\bibinfo {volume} {78}},\ \bibinfo
  {pages} {1396} (\bibinfo {year} {1997})}\BibitemShut {NoStop}%
\bibitem [{\citenamefont {Zanella}\ \emph {et~al.}(2008)\citenamefont
  {Zanella}, \citenamefont {Fagan}, \citenamefont {Mota},\ and\ \citenamefont
  {Fazzio}}]{zanella}%
  \BibitemOpen
  \bibfield  {author} {\bibinfo {author} {\bibfnamefont {I.}~\bibnamefont
  {Zanella}}, \bibinfo {author} {\bibfnamefont {S.~B.}\ \bibnamefont {Fagan}},
  \bibinfo {author} {\bibfnamefont {R.}~\bibnamefont {Mota}}, \ and\ \bibinfo
  {author} {\bibfnamefont {A.}~\bibnamefont {Fazzio}},\ }\href@noop {}
  {\bibfield  {journal} {\bibinfo  {journal} {The Journal of Physical Chemistry
  C}\ }\textbf {\bibinfo {volume} {112}},\ \bibinfo {pages} {9163} (\bibinfo
  {year} {2008})}\BibitemShut {NoStop}%
\bibitem [{\citenamefont {Valencia}\ \emph {et~al.}(2010)\citenamefont
  {Valencia}, \citenamefont {Gil},\ and\ \citenamefont {Frapper}}]{valencia}%
  \BibitemOpen
  \bibfield  {author} {\bibinfo {author} {\bibfnamefont {H.}~\bibnamefont
  {Valencia}}, \bibinfo {author} {\bibfnamefont {A.}~\bibnamefont {Gil}}, \
  and\ \bibinfo {author} {\bibfnamefont {G.}~\bibnamefont {Frapper}},\
  }\href@noop {} {\bibfield  {journal} {\bibinfo  {journal} {The Journal of
  Physical Chemistry C}\ }\textbf {\bibinfo {volume} {114}},\ \bibinfo {pages}
  {14141} (\bibinfo {year} {2010})}\BibitemShut {NoStop}%
\bibitem [{\citenamefont {Chan}\ \emph {et~al.}(2008)\citenamefont {Chan},
  \citenamefont {Neaton},\ and\ \citenamefont {Cohen}}]{chan-cohen}%
  \BibitemOpen
  \bibfield  {author} {\bibinfo {author} {\bibfnamefont {K.~T.}\ \bibnamefont
  {Chan}}, \bibinfo {author} {\bibfnamefont {J.~B.}\ \bibnamefont {Neaton}}, \
  and\ \bibinfo {author} {\bibfnamefont {M.~L.}\ \bibnamefont {Cohen}},\
  }\href@noop {} {\bibfield  {journal} {\bibinfo  {journal} {Phys. Rev. B}\
  }\textbf {\bibinfo {volume} {77}},\ \bibinfo {pages} {235430} (\bibinfo
  {year} {2008})}\BibitemShut {NoStop}%
\bibitem [{\citenamefont {Zhang}\ \emph {et~al.}(2012)\citenamefont {Zhang},
  \citenamefont {Lazo}, \citenamefont {Bluegel}, \citenamefont {Heinze},\ and\
  \citenamefont {Mokrousov}}]{stefan}%
  \BibitemOpen
  \bibfield  {author} {\bibinfo {author} {\bibfnamefont {H.}~\bibnamefont
  {Zhang}}, \bibinfo {author} {\bibfnamefont {C.}~\bibnamefont {Lazo}},
  \bibinfo {author} {\bibfnamefont {S.}~\bibnamefont {Bluegel}}, \bibinfo
  {author} {\bibfnamefont {S.}~\bibnamefont {Heinze}}, \ and\ \bibinfo {author}
  {\bibfnamefont {Y.}~\bibnamefont {Mokrousov}},\ }\href@noop {} {\bibfield
  {journal} {\bibinfo  {journal} {Phys. Rev. Lett.}\ }\textbf {\bibinfo
  {volume} {108}},\ \bibinfo {pages} {056802} (\bibinfo {year}
  {2012})}\BibitemShut {NoStop}%
\bibitem [{\citenamefont {Wang}(2008)}]{wang-sgs}%
  \BibitemOpen
  \bibfield  {author} {\bibinfo {author} {\bibfnamefont {X.~L.}\ \bibnamefont
  {Wang}},\ }\href@noop {} {\bibfield  {journal} {\bibinfo  {journal} {Phys.
  Rev. Lett.}\ }\textbf {\bibinfo {volume} {100}},\ \bibinfo {pages} {156404}
  (\bibinfo {year} {2008})}\BibitemShut {NoStop}%
\bibitem [{\citenamefont {Humphrey}\ \emph {et~al.}(1996)\citenamefont
  {Humphrey}, \citenamefont {Dalke},\ and\ \citenamefont {Schulten}}]{vmd}%
  \BibitemOpen
  \bibfield  {author} {\bibinfo {author} {\bibfnamefont {W.}~\bibnamefont
  {Humphrey}}, \bibinfo {author} {\bibfnamefont {A.}~\bibnamefont {Dalke}}, \
  and\ \bibinfo {author} {\bibfnamefont {K.}~\bibnamefont {Schulten}},\
  }\href@noop {} {\bibfield  {journal} {\bibinfo  {journal} {J. Molecular
  Graphics}\ }\textbf {\bibinfo {volume} {14}},\ \bibinfo {pages} {33}
  (\bibinfo {year} {1996})}\BibitemShut {NoStop}%
\bibitem [{\citenamefont {Kokalj}(2003)}]{xcrysden}%
  \BibitemOpen
  \bibfield  {author} {\bibinfo {author} {\bibfnamefont {A.}~\bibnamefont
  {Kokalj}},\ }\href@noop {} {\bibfield  {journal} {\bibinfo  {journal}
  {Computational Materials Science}\ }\textbf {\bibinfo {volume} {28}},\
  \bibinfo {pages} {155 } (\bibinfo {year} {2003})}\BibitemShut {NoStop}%
\end{thebibliography}
\end{document}